\documentclass[onecolumn,superscriptaddress,aps,jmp,amsmath,amssymb]{revtex4-1}

\usepackage{amsthm}
\usepackage{mathrsfs}
\usepackage{graphicx}
\usepackage{hyperref}
\usepackage{xcolor}
\usepackage{fourier}
\usepackage[normalem]{ulem}

\usepackage{caption}
\usepackage{subcaption}
\usepackage{color}
\usepackage{csquotes}
\usepackage{verbatim}
\usepackage{bookmark}
\usepackage{algorithmic} 
\usepackage{url,hyperref}
\usepackage{enumerate}
\usepackage{dsfont}

\allowdisplaybreaks

\theoremstyle{plain}
\newtheorem{theorem}{Theorem}
\newtheorem{lemma}{Lemma}

\newtheorem{properties}{Properties}[section]

\theoremstyle{definition}
\newtheorem{defn}[theorem]{Definition}

\theoremstyle{remark}
\newtheorem*{remark}{Remark}

\DeclareMathOperator{\pv}{p.v.}

\DeclareMathOperator{\spanop}{span}

\newcommand{\ie}{\textit{i.e.}}
\newcommand{\eg}{\textit{e.g.}}

\newcommand{\RR}{\mathbb{R}}

\newcommand{\CC}{\mathbb{C}}
\newcommand{\Or}{\mathcal{O}}

\newcommand{\Natural}{\mathbb{N}}
\newcommand{\Int}{\mathbb{Z}}

\newcommand{\Real}{\mathbb{R}}

\newcommand{\id}{\text{Id}}

\newcommand{\qinner}[2]{\langle#1 \vert #2\rangle}
\newcommand{\qInner}[2]{\left\langle#1 \vert #2\right\rangle}

\makeatletter
\newcommand*{\rom}[1]{\expandafter\@slowromancap\romannumeral #1@}
\makeatother

\IfFileExists{mathabx.sty}%
  {\DeclareFontFamily{U}{mathx}{\hyphenchar\font45}%
   \DeclareFontShape{U}{mathx}{m}{n}{<->mathx10}{}%
   \DeclareSymbolFont{mathx}{U}{mathx}{m}{n}%
   \DeclareFontSubstitution{U}{mathx}{m}{n}%
   \DeclareMathAccent{\widebar}{0}{mathx}{"73}%
}{%
  \PackageWarning{mathabx}{%
    Package mathabx not available, therefore\MessageBreak substituting
    widebar with overline\MessageBreak }%
  \newcommand{\widebar}[1]{\overline{#1}}%
}

\newcommand{\mc}[1]{\mathcal{#1}}

\newcommand{\eps}{\epsilon}

\newcommand{\norm}[1]{\lVert#1\rVert}

\newcommand{\bran}[1]{\langle#1}
\newcommand{\bra}[1]{\langle#1\rvert}
\newcommand{\Bra}[1]{\left\langle#1\right\rvert}
\newcommand{\ket}[1]{\lvert#1\rangle}
\newcommand{\Ket}[1]{\left\lvert#1\right\rangle}

\def\bigl{\mathopen\big}

\def\bigr{\mathclose\big}

\numberwithin{equation}{section}

\hypersetup{
    unicode=false,          
    pdftoolbar=true,        
    pdfmenubar=true,        
    pdffitwindow=false,     
    pdfstartview={FitH},    
    pdftitle={My title},    
    pdfauthor={Author},     
    pdfsubject={Subject},   
    pdfcreator={Creator},   
    pdfproducer={Producer}, 
    pdfkeywords={keyword1} {key2} {key3}, 
    pdfnewwindow=true,      
    colorlinks=true,       
    linkcolor=blue,          
    citecolor=blue,        
    filecolor=magenta,      
    urlcolor=blue           
}

\newcommand{\Anticom}[2]{\left[ #1, #2\right]_{+}}

\newcommand{\trb}{\text{Tr}_{b}}
\newcommand{\sysfreq}{\omega_s}
\newcommand{\rf}{Redfield}
\newcommand{\lb}{Lindblad}
\newcommand{\lbd}{Lindbladian}
\newcommand{\cme}{classical master equation}

\newcommand{\lcme}{\lbd\ \cme}
\newcommand{\sch}{Schr{\"o}dinger}
\newcommand{\ah}{Anderson-Holstein}
\newcommand{\nz}{Nakajima-Zwanzig}

\newcommand{\lo}{\mathcal{L}}
\newcommand{\po}{\mc{P}}
\newcommand{\qo}{\mc{Q}}
\newcommand{\lhm}{\hat{\mathcal{H}} }
\newcommand{\lhms}{\mathcal{H}}
\newcommand{\wig}{\varrho}

\newcommand{\rhoo}{\lambda}
\newcommand{\epe}{U} 
\newcommand{\eeng}{E}

\newcommand{\eengsc}{\mathsf{E}} 
\newcommand{\lensc}{\mathsf{\ell}}
\newcommand{\tsc}{\mathsf{T}}
\newcommand{\vsc}{\mathsf{V}}

\newcommand{\pert}{perturbation}
\newcommand{\pertt}{\pert\ theory}

\usepackage{setspace}
\begin{document}

\title{Lindblad equation and its semi-classical limit of the
  Anderson-Holstein model \thanks{This work is partially supported by the National
  Science Foundation under grants DMS-1454939.}}

\author{Yu Cao}\email{yucao@math.duke.edu}
\affiliation{Department of Mathematics, Duke University, Box 90320, Durham, NC 27708 USA}
\author{Jianfeng Lu}\email{jianfeng@math.duke.edu}
\affiliation{Department of Mathematics, Duke University, Box 90320, Durham, NC 27708 USA}
\affiliation{Department of Physics and Department of Chemistry, Duke University, Box 90320, Durham, NC 27708 USA}

\date{\today}

\begin{abstract}
  For multi-level open quantum system, the interaction between
  different levels could pose challenge to understand the quantum
  system both analytically and numerically.  In this work, we study
  the approximation of the dynamics of the Anderson-Holstein model, as
  a model of multi-level open quantum system, by \rf{} and \lb{}
  equations. Both equations have a desirable property that if the
  density operators for different levels is diagonal initially, they
  remain to be diagonal for any time. Thanks to this nice property,
  the semi-classical limit of both \rf{} and \lb{} equations could be
  derived explicitly; the resulting classical master equations share
  similar structures of transport and hopping terms.  The \rf{} and
  \lb{} equations are also compared from the angle of time dependent
  perturbation theory.
\end{abstract}

\keywords{Multi-level open quantum system, \ah{} model, \rf{} equation, \lb{} equation, semi-classical limit}

\maketitle

\section{Introduction}

Multi-level open quantum systems have received much attention due to
their wide applications and intriguing phenomena \cite{Ghosh04,
  Galperin07, Galperin08}.  One of the simplest models is perhaps the
\ah{} model, a two-level open quantum system \cite{holstein1959}. The
\ah{} model is a simplistic model for a molecule as the system of
interest, represented by a classical nucleus degree of freedom and a
two level electronic degree of freedom, coupled with a bath of
fermions, for instance, a reservoir of electrons. In this paper, the
simplified version of \ah{} model discussed in \cite{
  Wenjie15_friction} is used as an example for illustrating purpose,
which will be explained in more details in the next section. The goal
is to understand the approach of quantum master equations, in
particular, the \lbd\ formalism, for such systems in the
weak-coupling limit and also to study the semiclassical limit of the
quantum master equations.  Our study here should generalize to other
multi-level open quantum systems.

\ah{} model, since introduced, has been widely studied using various
theoretical and numerical approaches, for instance, the Green's
function approach \cite{Dmitry, Galperin04_Green}, equation-of-motion method
\cite{Galperin06_EOM, Galperin07_EOM}, quantum Monte Carlo method \cite{Rabani}, semi-classical approximation
\cite{Mitra04, Mitra05}, non-crossing approximation
\cite{ChenHsingTa2016} and by using quantum master equations
\cite{Elste2008, Esposito09, Esposito10, Wenjie15_friction,
  Wenjie15_ah, Wenjie15_broadening, Wenjie16_bcme}. In the perspective
of quantum master equation, which is mostly related to the current
work, the quantum master equation in Redfield flavor for \ah{} model has been derived in
\cite{Elste2008, Esposito09, Esposito10}. The semiclassical limit of
the \rf{} equation, known as the classical master equation (CME), has
been considered in \cite{Wenjie15_friction}, which also proposed a
numerical method based on surface hopping. The CME perspective has
then been used to study various physical aspects of Anderson-Holstein
model, e.g., broadening, Marcus rate \cite{Wenjie15_ah,
  Wenjie15_broadening, Wenjie16_bcme}. In all these works, the focus
has been on \rf{} equation (or \rf{} generator). As far as we know,
not much attention has been put into the \lbd\ formulation nor its
semiclassical limit of the Anderson-Holstein model, which is the focus
of the current work.

Recall that closed quantum systems can be fully characterized by the
Hamiltonian; its time-evolution dynamics is characterized by \sch\
equation (or von Neumann equation if we are dealing with density
operators). In the framework of quantum master equation, open quantum
systems can be described by \nz\ equation with the assistance of
projection operator to a subspace in which density operator for closed
system is separable \cite{breuer}.  Although \nz\ equation provides an
exact expression for the open quantum system in the interaction
picture, in general, it is not easy to retrieve useful information,
neither analytically nor numerically.  Part of the reason is
attributed to the memory effect in the \nz\ equation. While it is, of
course, important to research on non-Markovian dynamics itself; many
questions in non-Markovian dynamics are still open and the
mathematical foundation requires further investigation
\cite{breuer2016}. Often times Markovian approximation is taken to
simplify the governing equations, which is a valid approximation in
the weak coupling regime. See also \cite{Davies} for mathematical
study of the quantum Markovian approximation.

\begin{figure}[h!]
    \centering
    \includegraphics[width=0.90\textwidth]{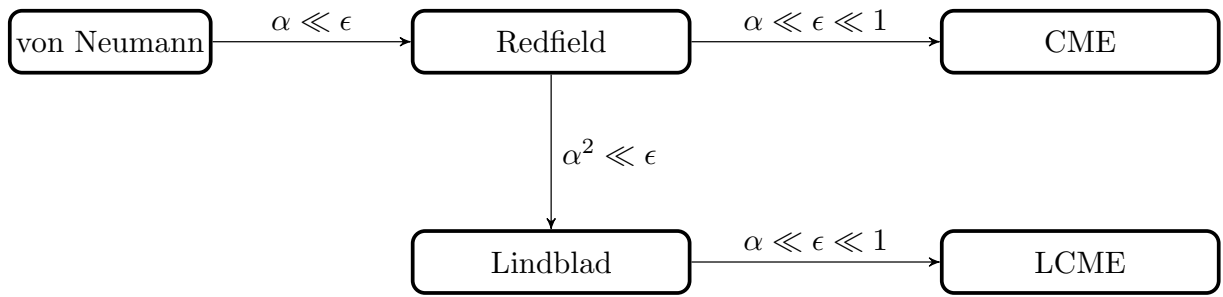}
    \caption{This diagram summarizes the conditions needed for
      approximation and connections between various
      models.  \label{fig::outline}}
\end{figure}

The Markovian approximation leads to \rf{} equation using
time-convolutionless (TCL) projection operator method in the
weak-coupling limit \cite{breuer}.  Furthermore, with secular
approximation, \lb{} equation can be obtained from \rf{} equation
\cite{breuer}; the Lindblad equation has better mathematical structures
such as complete positivity \cite{lindblad1976}.  The essence of
secular approximation is to approximate fast oscillating terms by zero
in the sense of averaging on a coarser time scale.  The goal of this
paper is to study \rf{} equation and \lb{} equation for multi-level
open quantum systems as well as their semi-classical limit, in the
context of the \ah{} model.  For these two equations, one obtains the
classical master equations (CME) and Lindbladian classical master
equations (LCME) in the semiclassical limit. The relations of various
models and the asymptotic limit connecting those are summarized in
Figure~\ref{fig::outline}.

It is worth mentioning that there is a debate in the literature on
which equation better models the open quantum system, especially when
the coupling between the system and the bath is not weak.  The
underlying discussion focuses on whether complete positivity (CP) is
necessary for modeling open quantum systems.  There are at least two
arguments supporting complete positivity in quantum systems: the first
one is from the perspective of \enquote{total domain}; the second one
from \enquote{product state} \cite{McCracken2013}. What's more, one
recent research indicates that without complete positivity in \rf{}
equation, the dynamics might be inconsistent with the second law of
thermodynamics \cite{argentieri}.  Some, however, criticize that we
might over-emphasize the importance of complete positivity in modeling
open quantum systems. Pechukas proposed that for a composite quantum
system with entangled initial condition, the positivity property might not hold for the reduced dynamics \cite{Pechukas1994}. Shaji and Sudarshan argued
that complete positivity is not necessary by carefully examinizing
arguments supporting complete positivity
\cite{Shaji200548}. Negativity, as opposed to positivity, is not only
observed in experiment but also can be informative to the coupling
with bath \cite{McCracken2013}. In our study on Anderson-Holstein
model, imposing complete positivity (and thus \lb{} equation) should
be justified as we only consider the weak-coupling regime.  In
particular, as will become clear in our analysis, under the same
assumption used in deriving Redfield equation, the secular
approximation for getting Lindblad equation is in fact also justified;
hence, the use of Lindblad equation is natural.

In this paper, we consider \ah{} model in weak-coupling and
semi-classical limits.  Under the assumption that the coupling
strength is weak, we will revisit the derivation of \rf{} equation in
Section \ref{sec::rf_equation} and derive the explicit form of \lb{}
equation in Section \ref{sec::lindblad}. The semi-classical study of
both equations is discussed in Section \ref{sec::semi-classical}. The
\pert\ result of both equations is presented in Section
\ref{sec::time-dependent}. Section \ref{sec::conclusion} summarizes
the main results and ends the paper with some concluding remarks.

\section{Anderson-Holstein model}
\label{sec::model}

The Anderson-Holstein model under study here describes a two-level system
coupled with a bath of many non-interacting electrons (or in general
spin-$1/2$ fermions). For instance, the two-level system can be
thought as a simplistic model for the nuclei degree of freedom of a
molecule with two potential energy surfaces depending on the
electronic state of the molecule. For simplicity, in the
Anderson-Holstein model, the two-level system is in one spatial
dimension and one of the potential energy surface is taken to be a
harmonic oscillator with frequency $\sysfreq$, and the difference
$\epe(x)$ between the two potential energy surface is modeled as a
linear function of the nucleus position (and thus is also a harmonic
oscillator with shifted center and energy) \cite{holstein1959}.  More
specifically, the Hamiltonian for the whole system is given by
\begin{equation}\label{eq:totalham}
\begin{split}
\hat{H} &= \hat{H}_s + \hat{H}_b + \hat{H}_c \\
\hat{H}_s &=  \frac{\hat{p}^2}{2m} + \frac{1}{2}m\sysfreq^2 \hat{x}^2 + \epe(\hat{x}) \hat{d}^{\dagger} \hat{d}\\
\hat{H}_b &= \sum_{k} (\eeng_k - \mu)\hat{c}_k^{\dagger} \hat{c}_k \\
\hat{H}_c &= \sum_{k} V_k (\hat{c}_k^{\dagger} \hat{d} + \hat{c}_k \hat{d}^{\dagger}  ) \\
\end{split}
\end{equation}
where we follow the notation of \cite{Wenjie15_friction}: $\hat{d}$
and $\hat{d}^{\dagger}$ are the annihilation and creation operators
for the two-level electron state of the molecule, $\hat{c}_k$ and
$\hat{c}^{\dagger}_k$ are the annihilation and creation operators for
electron states in the bath, $\eeng_k$ is the energy level of those
states, $\mu$ is the Fermi level, and $V_k$ is the coupling strength
between the molecule and the $k$-th mode in bath, assumed to be
real. The Hilbert space corresponds to the molecule is thus
$L^2(\RR) \otimes \CC^2 = L^2(\RR) \otimes \spanop\bigl\{ \ket{0},
\ket{1}\bigr\}$,
so that we have $\hat{d}\ket{1} = \ket{0}$ and
$\hat{d}^{\dagger} \ket{0} = \ket{1}$.  Thus in \eqref{eq:totalham},
$\hat{H}_s$ is the Hamiltonian operator of the ``system'', $\hat{H}_b$
is the Hamiltonian of the ``bath'', and $\hat{H}_c$ describes the
coupling between the system and the bath. The goal is to understand
the evolution of the system as an open quantum system (\textit{i.e.},
integrate out the bath degree of freedom). 
The above is a simplified version of the original \ah{}
  model as (1) only one electron is considered for the molecule, thus
  the model excludes Coulomb interaction between electrons of the
  molecule; (2) the molecule is coupled to one electrode, but not two
  electrodes, and thus only one Fermi level $\mu$ is used for the
  environment.

\begin{remark}
  In the literature, \eg{} \cite{Mitra05}, sometimes the term
  \enquote{single-level} is used for the above system to emphasize
  that there is only one on-site electron. We have adopted here the
  term \enquote{two-level} to emphasize that the Hilbert space for
  molecular system is
  $L^2(\Real)\otimes \text{span}\bigl\{ \ket{0}, \ket{1} \bigr\}$ and the second
  component has dimension $2$. 
\end{remark}

To proceed, let us first non-dimensionalize the problem
according to the following rescaling:
\begin{enumerate}
\item Denote $\lensc$ the characteristic length scale of $x$, the
  position degree of freedom of the system. That is, if we take
  $x = \lensc\tilde{x}$, $\tilde{x}$ becomes a dimensionless quantity
  with order $\mathcal{O}(1)$.
\item As a consequence, the scaling factor for molecular energy is
  then $\eengsc = m \sysfreq^2 \lensc^2$. We will use $\eengsc$ as
  the scaling factor for all physical quantities whose dimension is
  energy (thus including all terms in the Hamiltonian). 
\item Denote $\tsc$ the time scale of the evolution of the system. Thus,
  $t = \tsc \tilde{t}$ where $\tilde{t} = \mathcal{O}(1)$. Physically, it
  is reasonable to choose $\tsc = \frac{1}{\sysfreq}$ \cite{shankar},
  since this is the time scale of an isolated harmonic oscillator
  with frequency $\sysfreq$.
\item As both $\hbar\sysfreq$ and $\eengsc$ have the energy dimension,
  the ratio
  \begin{equation*}
    \epsilon := \frac{\hbar\sysfreq}{\eengsc}
  \end{equation*}
  is a dimensionless quantity. In our analysis of the semi-classical
  limit of the system, we will assume that $\epsilon$ is a small
  parameter $\epsilon\downarrow 0$.

\item Let $\vsc$ be the typical interaction strength with dimension as
  energy, \textit{i.e.}, we assume
  $\tilde{V}_k := \frac{V_k}{\vsc} = \mathcal{O}(1)$. The ratio of $\vsc$ and $\eengsc$
  \begin{equation*}
    \alpha := \frac{\vsc}{\eengsc}
  \end{equation*} 
  is dimensionless. Weak-coupling limit means
  $\vsc \ll \eengsc$, and hence $\alpha \downarrow 0$, which means
  physically that the coupling between the bath and the system is weak
  compared to the typical energy scale of the system. Note that by the
  above rescaling, we have
  \[\frac{V_k}{\eengsc} = \alpha \tilde{V}_k.\] 
\end{enumerate}
After performing the above rescaling, the non-dimensionalized
Hamiltonian becomes
\begin{align*}
  \hat{H}_{non} &= \hat{H}_{s,non} + \hat{H}_{b,non} + \alpha \hat{H}_{c,non} \\
  \hat{H}_{s,non} &=  - \frac{1}{2}\epsilon^2  \nabla_{\tilde{x}}^2 + \frac{1}{2}\tilde{x}^2 + \tilde{\epe}(\tilde{x}) \hat{d}^{\dagger} \hat{d}\\
  \hat{H}_{b,non} &= \sum_{k} (\tilde{\eeng}_k - \tilde{\mu})\hat{c}_k^{\dagger} \hat{c}_k \\
  \hat{H}_{c,non} &=  \sum_{k} \tilde{V}_k (\hat{c}_k^{\dagger} \hat{d} + \hat{c}_k \hat{d}^{\dagger}  ) 
\end{align*}
where $\tilde{\epe}(\tilde{x}) = \frac{\epe(x)}{\eengsc}$, $\tilde{\eeng}_k = \frac{\eeng_k}{\eengsc}$ and $\tilde{\mu} = \frac{\mu}{\eengsc}$.
Moreover, the von Neumann equation becomes
\[i \epsilon \partial_{\tilde{t}} \hat{\rho} = \bigl[ \hat{H}_{non}, \hat{\rho}\bigl],\]
where $\hat{\rho}$ is the density operator for the closed system and
$\bigl[\ ,\ \bigl]$ is the usual commutator.

Dropping \enquote{tilde} and \enquote{non} and shifting the energy
reference to replace $E_k -\mu$ by $E_k$ to simplify the notation, we
arrive at the Hamiltonian
\begin{equation}
\label{eqn::hamiltonian}
\begin{split}
\hat{H} &= \hat{H}_{s} + \hat{H}_{b} + \alpha \hat{H}_{c} \\
\hat{H}_{s} &=  - \frac{1}{2}\epsilon^2  \nabla_{x}^2 + \frac{1}{2} x^2 + \epe(x) \hat{d}^{\dagger} \hat{d}\\
\hat{H}_{b} &= \sum_{k} 
\eeng_k
\hat{c}_k^{\dagger} \hat{c}_k \\
\hat{H}_{c} &=  \sum_{k} V_k (\hat{c}_k^{\dagger} \hat{d} + \hat{c}_k \hat{d}^{\dagger}  ) =:  \hat{C}^{\dagger} \hat{d} + \hat{C} \hat{d}^{\dagger}\\
\end{split}
\end{equation}
Here, $U(x)$ is a linear function of position $x$ with the form (the reason of the specific choice or parametrization will become clear below)
\[
\epe(x) := \sqrt{2} g x + g^2 + \bar{\epe}_0,
\]
where $\bar{\epe}_0$ is known as renormalized energy, and 
the weighted annihilation operator $\hat{C}$ is defined as
\[
\hat{C} := \sum_{k} V_k \hat{c}_k.
\]
The evolution of the density operator is given by the von Neumann equation 
\begin{equation}
\label{eqn::von_Neumann}
i \epsilon \partial_{t} \hat{\rho} = \bigl[ \hat{H}, \hat{\rho}\bigl].
\end{equation}
All quantities in the Equations \eqref{eqn::hamiltonian} and
\eqref{eqn::von_Neumann} are dimensionless and the parameters
$ V_k$ and $\eeng_k$'s are $\Or(1)$. 

In summary, after non-dimensionalization, the model contains two
scaling parameters $\eps$ and $\alpha$, corresponding to the
semiclassical parameter and coupling strength respectively. In the
rest of the paper, we will consider the weak-coupling limit and the
semiclassical limit. In the weak-coupling limit, we have
$\alpha \downarrow 0$, which leads to \rf{} and \lb{} equations for
fixed $\eps$, while the semiclassical limit means
$\epsilon \downarrow 0$. See Figure~\ref{fig::outline} for an overview.

Notice that 
\begin{equation*}
  \hat{H}_0 := \bra{0} \hat{H}_s \ket{0} \equiv  -\frac{1}{2}\epsilon^2 \nabla_{x}^2 + \frac{1}{2}x^2
\end{equation*}
is a Hamiltonian for a single harmonic oscillator.  It is well-known
that it has eigenfunctions 
\begin{equation*}
  \phi^0_{k} = N_k^0
  H_k\left(\frac{x}{\sqrt{\epsilon}}\right)\exp\left(-\frac{x^2}{2\epsilon}\right), \qquad k \in \Natural,
\end{equation*}
where $H_k$ is the $k$-th Hermite polynomial 
and $N_k^0$ is a normalization constant. The corresponding eigenvalue
is $\epsilon(k+\frac{1}{2})$. Similarly,
\begin{equation*}
  \hat{H}_1 := \bra{1} \hat{H}_s \ket{1} \equiv -\frac{1}{2}\epsilon^2
  \nabla_{x}^2 + \frac{1}{2}(x+ \sqrt{2} g)^2 + \bar{\epe}_0
\end{equation*}
has eigenfunctions $\phi^1_{k} = \phi_k^0(x+\sqrt{2} g)$ and
corresponding eigenvalues $\epsilon (k+\frac{1}{2}) +
\bar{\epe}_0$.
Note that this explains the specific parametrization of the linear
function $U(x)$ above.  To keep notations and calculations simple, we
will only consider the \ah{} model with renormalized energy
$\bar{\epe}_0 = 0$. The extension to the general case is straightforward.

\section{Revisiting the Derivation of \rf{} Equation}
\label{sec::rf_equation}

The derivation of \rf{} equation has been well studied and presented
in e.g., \cite{breuer, densitymat, Elste2008}. Physically,
Born-Markov approximation is the key to reduce the dynamics of the
system to a Markovian dynamics. In this section, we will revisit the
derivation for the \ah{} model to set the grounds of our discussion
below, using the time-convolutionless equation (TCL) approach
following \cite[Chapter 9]{breuer}; we will borrow notations from
this reference as well. For fixed $\epsilon$, we consider the
weak-coupling
limit below, that is,
$\alpha\downarrow 0$ while $\epsilon$ stays fixed, and hence 
$\eta := \frac{\alpha}{\epsilon}\downarrow 0$.

To simplify the dynamical equation, it is more convenient to use the
interaction picture (with respect to the uncoupled system and bath),
so that the operators are given by
\begin{equation*}
  \hat{O}_{I}(t) := e^{\frac{i}{\epsilon} (\hat{H}_s +\hat{H}_b) t}
\hat{O} e^{-\frac{ i}{\epsilon} (\hat{H}_s +\hat{H}_b) t}.
\end{equation*}
Interaction picture is very convenient in weak-coupling limit since it
removes the effect of fast motion (due to $\hat{H}_s+\hat{H}_b$) from
the slow motion (due to $\alpha \hat{H}_c$).

The von Neumann equation in the interaction picture is
\begin{equation}
\label{eqn::von-neu}
\frac{d}{dt}\hat{\rho}_{I}(t) = - i \frac{\alpha}{\epsilon} \bigl[\hat{H}_{c,I}(t), \hat{\rho}_{I}(t) \bigl] =: \eta \lo(t) \hat{\rho}_I(t)
\end{equation}
where we have introduced
$\lo(t) := -i \bigl[\hat{H}_{c,I}(t), \cdot \bigl]$ as a
super-operator acting on density operators. 
By explicit calculation, we have 
\begin{equation*}
\hat{H}_{c,I}(t) = \hat{C}^{\dagger}_{I}(t) \hat{d}_{I}(t) +
\hat{d}^{\dagger}_{I}(t) \hat{C}_{I}(t),
\end{equation*}
 where
\[ \hat{d}_{I}(t) = e^{\frac{i}{\epsilon} \hat{H}_s t} \hat{d} e^{-\frac{i}{\epsilon} \hat{H}_s t }, \qquad \text{and} \qquad \hat{C}_{I}(t) = \sum_{k} V_k e^{-\frac{i}{\epsilon} \eeng_k t} \hat{c}_k.\]

For any trace-class operator $A$ defined for the whole closed system, 
we define a projection operator $\po$ by 
\[\po A = \trb (A) \otimes \hat{\rho}_{b,eq} \]
where $\trb(A)$ is the partial trace over bath degree of freedom, and
where $\hat{\rho}_{b,eq} := e^{-\beta \hat{H}_b} /Z_b$ is the density
operator of electron bath at thermal equilibrium, with
$\beta = \frac{\eengsc}{k_{B} T}$ (inverse of the rescaled
temperature) and $Z_b$ is the partition function. This projection
operator $\po$ disentangles the system and bath and replaces the bath
by the thermal equilibrium; this is a core ingredient in Born
approximation, whose physical reasoning can be found in
\cite[p~276]{densitymat}. We also define its orthogonal complement as
$\qo := \id - \po$. For a given density matrix $\hat{\rho}_{I}(t)$,
$\po \hat{\rho}_{I}(t)$ is known as relevant part and
$\qo \hat{\rho}_{I}(t)$ irrelevant part.

We may formally write down the solution to  Equation \eqref{eqn::von-neu} using Green's function as 
\begin{equation}\label{eq:solvonneu}
  \hat{\rho}_{I}(s) = G(t,s) \hat{\rho}_{I}(t) = G(t,s) (\po+\qo) \hat{\rho}_{I}(t),\qquad (\text{for } s \le t)
\end{equation}
where $G(t,s) := T_{\rightarrow} \exp\left( -\eta \int_{s}^{t} ds' \lo(s') \right)$ and $T_{\rightarrow}$ represents anti-chronological time-ordering operator. 
On the other hand, applying operator $\qo$ to Equation
\eqref{eqn::von-neu} gives a differential equation
\[
\frac{d}{dt} \qo \hat{\rho}_{I}(t) = \eta \qo \lo(t) \po
\hat{\rho}_{I}(t) + \eta \qo \lo(t) \qo \hat{\rho}_{I}(t). 
\] 
By
Duhamel's principle, its solution in integral form is
\[ \begin{split}
\qo \hat{\rho}_{I}(t) &= \mathcal{G}(t, t_0) \qo \hat{\rho}_{I}(t_0) + 
\eta \int_{t_0}^{t}ds\ \mathcal{G}(t,s) \qo \lo(s) \po \hat{\rho}_I(s) \\
&\stackrel{\eqref{eq:solvonneu}}{=} \mathcal{G}(t, t_0) \qo \hat{\rho}_{I}(t_0) + 
\eta \int_{t_0}^{t}ds\ \mathcal{G}(t,s) \qo \lo(s) \po G(t,s) (\po+\qo) \hat{\rho}_{I}(t) \\
&= \mathcal{G}(t, t_0) \qo \hat{\rho}_{I}(t_0) + 
\Sigma(t) (\po+\qo) \hat{\rho}_{I}(t)
\end{split}\]
where $t_0$ is the starting time of interest, $\mathcal{G}(t,s) = T_{\leftarrow} \exp\left(   \eta \int_{s}^{t}ds'\  \qo \lo(s') \right)$,
$T_{\leftarrow}$ represents the 
chronological time-ordering operator,
and $\Sigma(t) := \eta \int_{t_0}^{t}ds\ \mathcal{G}(t,s) \qo \lo(s) \po G(t,s)
$. 
Assume that at time $t_0$, the bath is at thermal equilibrium and density operator $\hat{\rho}_{I}(t_0)$ is separable, i.e., $\hat{\rho}_{I}(t_0) = \hat{\rho}_{s,I}(t_0)\otimes \hat{\rho}_{b,eq}$. Then $\qo \hat{\rho}_{I}(t_0) = 0$ and hence we obtain
\begin{equation*}
  \qo \hat{\rho}_{I}(t) = \left(\id - \Sigma(t)\right)^{-1} \Sigma(t) \po \hat{\rho}_{I}(t)
\end{equation*}
if $\id - \Sigma(t)$ is invertible, which is the case, for instance, when $\eta$ is so small  that $\norm{\Sigma(t)} < 1$. 

We may also apply the operator $\po$ to Equation \eqref{eqn::von-neu}
and get
\[ \begin{split}
\frac{d}{dt} \po \hat{\rho}_{I}(t) & = \eta \po \lo(t) \po \hat{\rho}_{I}(t) + \eta \po \lo(t) \qo \hat{\rho}_{I}(t)\\
&= \eta \po \lo(t) \po \hat{\rho}_{I}(t) + \eta \po \lo(t) (\id - \Sigma(t))^{-1} \Sigma(t) \po \hat{\rho}_{I}(t)\\
&= \eta \po \lo(t)  (\id - \Sigma(t))^{-1} \po \hat{\rho}_{I}(t) \\
& = \left(\eta  \po \lo(t) \po + \eta^{2} \po \lo(t) \int_{t_0}^{t}ds\ \qo \lo(s) \po \right)\hat{\rho}_{I}(t) + \mathcal{O}(\eta^3)
 \end{split} \]
In the last step we perform asymptotic expansion of operator in terms of $\eta$. It could be easily verified that $\po \lo(t) \po = 0$ from the definition of $\lo$. Then the leading order expansion is 
\[ \begin{split}
\frac{d}{dt} \po \hat{\rho}_{I}(t) &=  
\eta^2 \int_{t_0}^{t} ds\ \po \lo(t) \lo(s) \po \hat{\rho}_{I}(t) \\
\end{split}\]
After replacing $\lo$ by its definition, we arrive at 
\begin{equation}
\frac{d}{dt} \hat{\rho}_{s,I}(t) = -\left(\frac{\alpha}{\epsilon}\right)^2  \trb \left(  \int_{t_0}^{t}\ ds  \bigl[  \hat{H}_{c,I}(t), [ \hat{H}_{c,I}(s), \hat{\rho}_{s,I}(t)\otimes \hat{\rho}_{b,eq}]   \bigl] \right).
\end{equation}
The leading order expansion of $\frac{d}{dt}\hat{\rho}_{s,I}(t)$ is
the same as applying Born-Markov approximation to von Neumann equation
directly.
Change the variable $\tau = t - s$ and push $t_0$ to approach $-\infty$, then the last equation becomes \emph{Redfield equation}
\begin{equation}
\label{eqn::redfield}
\frac{d}{dt}\hat{\rho}_{s,I}(t) =- \left(\frac{\alpha}{\epsilon} \right)^2 \trb \left( \int_{0}^{\infty} d\tau \ \left[\hat{H}_{c,I}(t), [\hat{H}_{c,I}(t-\tau), \hat{\rho}_{s,I}(t) \otimes \hat{\rho}_{b,eq} ]\right]\right)
\end{equation}
There are two formal justifications for pushing $t_0$ to $-\infty$: if $t_0 = -\infty$, the dynamics does not depend on initial time as a parameter; moreover, if the system evolves from long time ago, we may as well consider $t_0 = -\infty$.

After opening the double commutator and simplify the equation, we arrive at
\begin{equation}
\label{eqn::redfield_2}
\begin{split}
\frac{d}{dt}\hat{\rho}_{s,I}(t) = 
- \left(\frac{\alpha}{\epsilon} \right)^2 \int_{0}^{\infty} d\tau\ 
& \bigl[\hat{d}_I(t), \hat{d}^{\dagger}_I(t-\tau) \hat{\rho}_{s,I}(t)\bigl] F(t, t-\tau) \\
+ & \bigl[\hat{\rho}_{s,I}(t) \hat{d}^{\dagger}_{I}(t-\tau), \hat{d}_I(t)\bigl]  G(t,t-\tau) + h.c. \\
\end{split}
\end{equation}
where time correlation functions
\begin{equation}
\begin{split}
F(t, t') &:= \trb\left( \hat{C}_{I}^{\dagger}(t) \hat{C}_{I}(t') \hat{\rho}_{b,eq}\right) = \sum_{k} V_k^2 \exp\left(\frac{i}{\epsilon} \eeng_k(t-t')\right) f(\eeng_k)\\
G(t, t') &:= \trb \left(\hat{C}_I(t') \hat{C}_{I}^{\dagger}(t) \hat{\rho}_{b,eq}\right) = \sum_{k} V_k^2 \exp\left(\frac{i}{\epsilon} \eeng_k(t-t')\right)(1-f(\eeng_k))\\
\end{split}
\end{equation}
with $f(z)$ being the Fermi-Dirac function 
$f(z) = 1/(1+e^{\beta z})$.

Transforming Equation \eqref{eqn::redfield_2} back into Schr{\"o}dinger picture,
we end up with 
\begin{equation}
\label{eqn::redfield_sch}
\begin{split}
\frac{d}{dt}\hat{\rho}_s(t) = -\frac{i}{\epsilon} \bigl[\hat{H}_s, \hat{\rho}_s(t)\bigl] 
- \left(\frac{\alpha}{\epsilon} \right)^2 \int_{0}^{\infty}d\tau\ 
& \left( \hat{d} e^{-\frac{i}{\epsilon}\hat{H}_s\tau } \hat{d}^{\dagger} e^{\frac{i}{\epsilon}
\hat{H}_s \tau } \hat{\rho}_s(t) - 
e^{-\frac{i}{\epsilon}\hat{H}_s \tau } \hat{d}^{\dagger} e^{\frac{i}{\epsilon} \hat{H}_s \tau} 
\hat{\rho}_s(t)\hat{d}\right) F(t, t-\tau) \\
+& \left( \hat{\rho}_s(t) e^{-\frac{i}{\epsilon}\hat{H}_s \tau}\hat{d}^{\dagger} e^{\frac{i}{\epsilon} \hat{H}_s \tau} \hat{d} - \hat{d} \hat{\rho}_s(t) e^{-\frac{i}{\epsilon} \hat{H}_s\tau} \hat{d}^{\dagger} e^{\frac{i}{\epsilon}\hat{H}_s\tau}   \right) G(t, t-\tau) \\
+ & h.c.\\
\end{split}
\end{equation} 
This is the \rf{} equation for Anderson-Holstein model.

\section{Derivation of \lb{} Equation}
\label{sec::lindblad}

It is a fundamental result \cite{lindblad1976} that a completely
positive dynamical map can be written in Lindblad form for open
quantum systems.  Thus, in this section, we aim at deriving the \lb\
equation for \ah\ model.  The derivation of Lindblad equation from a
microscopic point of view has been studied for some cases, see e.g.,
\cite{breuer}, though to the best of our knowledge not for the \ah\
model.  In this section, we will show that under the previous
condition that
\[\alpha\ll \eps, \quad \alpha \ll 1\]
\lb{} equation in \sch{} picture for the \ah{} model is given by 
\begin{equation}
\frac{d}{dt}\hat{\rho}_s(t) = -\frac{i}{\epsilon} \bigl[\hat{H}_s + \alpha^2 \lhm, \hat{\rho}_s(t)\bigl] + \frac{\alpha^2}{\epsilon}\mathcal{D}(\hat{\rho}_s(t))
\end{equation}
with a \emph{\lbd{} corrected Hamiltonian}
\begin{equation}
\begin{split}
\lhm  = \sum_{\omega\in \Int} b_{F}(\omega) \hat{D}(\omega)\hat{D}^{\dagger}(\omega) - 
b_{G}(\omega) \hat{D}^{\dagger}(\omega) \hat{D}(\omega)
\end{split}
\end{equation}
and dissipative operator
\begin{equation}
\begin{split}
\mathcal{D}(\hat{\rho}_{s}(t)) = \sum_{\omega\in \Int} 
& a_{F}(\omega) \left(\hat{D}^{\dagger}(\omega) \hat{\rho}_{s}(t) \hat{D}(\omega) - \frac{1}{2} 
\Anticom{
\hat{D}(\omega) \hat{D}^{\dagger}(\omega)} {\hat{\rho}_{s}(t)} 
\right) \\
+ & a_{G}(\omega) \left(\hat{D}(\omega) \hat{\rho}_{s}(t) \hat{D}^{\dagger}(\omega) - \frac{1}{2}
\Anticom{\hat{D}^{\dagger}(\omega) \hat{D}(\omega)}
{ \hat{\rho}_{s}(t) }
\right),
\end{split}
\end{equation}
where $\Anticom{\hat{A} }{\hat{B}}  := \hat{A}\hat{B} + \hat{B}\hat{A}$ is anti-commutator for two operators $\hat{A}$ and $\hat{B}$.
The coefficients will be given in Equation \eqref{eqn::coefs} below
and the operators $\hat{D}^{(\dagger)}(\omega)$ will be defined below
(see Equation \eqref{eqn::defn_D}). It is clear that the above
dissipative operator takes the Lindblad form.

\subsection{An alternative representation for \rf{} equation}
In Anderson-Holstein model, it is natural to 
consider eigenfunctions of $\hat{H}_s$,
which form two energy ladders. 
The evolution of system can be considered as quantum jumping between different energy levels. 
Thus the annihilation and creation operators 
might be decomposed in terms of the numbers of energy levels that the system jumps. Such decomposition has been used to derive \lb\ equation in \cite[p~125-131]{breuer}. We shall use this technique to study \rf\ equation and \lb\ equation below for \ah\ model.
\begin{defn}
For each $\omega \in \Int$, define an operator
\begin{equation}
\label{eqn::defn_D}
\hat{D}(\omega) := \sum_{k' - k = \omega} \Pi^{(0)}_{k}\ \hat{d}\ \Pi^{(1)}_{k'}
\end{equation}
where $\Pi^{(m)}_{k}$ is the projection operator to quantum state $\ket{\phi^{m}_k}\otimes \ket{m} \equiv \ket{\phi_{k}^m, m}$, $m \in \{0,1\}$, $k, k'\in \Natural$. 
\end{defn}

Recall that $\ket{\phi_{k}^{m}}$ is the eigenfunction of Hamiltonian $\hat{H}_s$ discussed at the end  of Section \ref{sec::model}.
In other words, 
$\Pi_{k}^{(m)} := \ket{\phi_{k}^m, m}\bra{\phi_{k}^m, m}$. 
Hence, the adjoint operator of $\hat{D}(\omega)$ is 
\begin{equation}
\hat{D}^{\dagger}(\omega)
= \sum_{k' - k = \omega} 
\Pi^{(1)}_{k'}\ \hat{d}^{\dagger}\ \Pi^{(0)}_{k}
\end{equation}

This definition was used in \cite{breuer} for a slightly different form of coupling Hamiltonian but it is also applicable here in \ah{} model. It can also be checked that properties proposed in \cite{breuer} still hold:
\begin{properties}
\normalfont
\begin{enumerate}[(I)]
\item $\hat{D}(\omega)$ and $\hat{D}^{\dagger}(\omega)$ are eigen-operators of $\hat{H}_s$, namely,
\begin{equation}
\bigl[\hat{H}_s, \hat{D}(\omega)\bigl] = -\epsilon \omega \hat{D}(\omega)\qquad
\bigl[\hat{H}_s, \hat{D}^{\dagger}(\omega)\bigl] = \epsilon \omega \hat{D}^{\dagger}(\omega)
\end{equation}

\item In the interaction picture, $\hat{D}_{I}(\omega,t)$ and $\hat{D}^{\dagger}_{I}(\omega,t)$
has the form
\begin{equation}
\label{eqn::D_sch_int}
\begin{split}
\hat{D}_{I}(\omega,t) \equiv & e^{\frac{i }{\epsilon}\hat{H}_s t} \hat{D}(\omega) e^{-\frac{i }{\epsilon} \hat{H}_s t}
= e^{-i\omega t} \hat{D}(\omega)\\
\hat{D}^{\dagger}_{I}(\omega,t) \equiv & e^{ \frac{i }{\epsilon}\hat{H}_s t} \hat{D}^{\dagger}(\omega) e^{-\frac{i }{\epsilon}\hat{H}_s t} 
= e^{i\omega t} \hat{D}^{\dagger}(\omega)\\
\end{split}
\end{equation}

\item $\hat{d}$ and $\hat{d}^{\dagger}$ can be decomposed into $\hat{D}(\omega)$ and $\hat{D}^{\dagger}(\omega)$ respectively. More specifically,
\begin{equation}
\label{eqn::d_decmp}
\hat{d} = \sum_{\omega} \hat{D}(\omega)\qquad 
\hat{d}^{\dagger} = \sum_{\omega} \hat{D}^{\dagger} (\omega)
\end{equation}

\item Then we can decompose coupling Hamiltonian $\hat{H}_c$ as
\begin{equation}
\hat{H}_c = \sum_{\omega} \left( \hat{D}(\omega) \otimes \hat{C}^{\dagger}  + \hat{D}^{\dagger}(\omega) \otimes \hat{C}\right)
\end{equation}
and in the interaction picture
\begin{equation}\label{eqn::ah_H_c_secular}
\hat{H}_{c,I} (t) = \sum_{\omega} \left( e^{-i\omega t} \hat{D}(\omega) \otimes \hat{C}_{I}^{\dagger}(t) + e^{i\omega t} \hat{D}^{\dagger}(\omega) \otimes \hat{C}_{I}(t) \right)
\end{equation}
\end{enumerate}
\end{properties}

These results directly follow from definition of $\hat{D}(\omega)$. Equation \eqref{eqn::d_decmp} is essential in decomposing $\hat{d} (\hat{d}^{\dagger})$ in terms of levels of jumping. The reason that $\omega$ is the level of jumping can be observed from the definition that $\hat{D}(\omega)$ maps quantum state $\ket{\phi_{k'}^{1},1}$ to quantum state $\ket{\phi_{k}^0,0}$ where $k = k' -\omega $. 
To prove the decomposition of $\hat{d}$ in terms of $\hat{D}(\omega)$, we use the completion relation $\sum_{k} (\Pi^{(0)}_{k} + \Pi^{(1)}_{k}) = \id$,
\[\begin{split}
\hat{d} &= \sum_{k, k'} \left(\Pi^{(0)}_{k} + \Pi^{(1)}_{k} \right)\ \hat{d}\ \left(\Pi^{(0)}_{k'} + \Pi^{(1)}_{k'}  \right)
= \sum_{k, k'} \Pi^{(0)}_{k}\ \hat{d}\ \Pi^{(1)}_{k'} \\
&= \sum_{\omega} \sum_{k' - k = \omega} \Pi^{(0)}_{k}\ \hat{d}\ \Pi^{(1)}_{k'} 
= \sum_{\omega} \hat{D}(\omega) \\
\end{split}\]
In the second step, we have used
$\hat{d} \ket{\phi_{k'}^{0}, 0} = \hat{d}^{\dagger} \ket{\phi_{k}^1,
  1} = 0$
for any $k, k'\in \Natural$.  It follows after taking the Hermitian
conjugate that $\bra{\phi_{k}^1, 1} \hat{d} = 0$; that is why there is
only one term $\Pi^{(0)}_{k}\ \hat{d}\ \Pi^{(1)}_{k'}$ left.  In the
third step, re-order the double summation is employed to first sum
over all differences of levels, namely, $\omega$ and then sum over all
possible combination of $k', k \in \Natural$ such that
$k' - k = \omega$, where the latter sum gives $\hat{D}(\omega)$.

Starting from Equation \eqref{eqn::redfield}, replacing $\hat{H}_{c,I}$ by Equation \eqref{eqn::ah_H_c_secular} and opening the double commutators, we arrive at an alternative representation of \rf{} equation
\begin{equation}
\label{eqn::rf_opd}
\begin{split}
\frac{d}{dt}\hat{\rho}_{s,I}(t) = - \frac{\alpha^2}{\epsilon} \sum_{\omega,\ \omega'}\ 
& e^{-i(\omega'-\omega) t}  \left(\hat{D}(\omega') \hat{D}^{\dagger}(\omega) \hat{\rho}_{s,I}(t) - \hat{D}^{\dagger}(\omega) \hat{\rho}_{s,I} (t) \hat{D}(\omega') \right)F(\omega) \\
+ & e^{-i(\omega'-\omega) t} \left(\hat{\rho}_{s,I}(t) \hat{D}^{\dagger}(\omega) \hat{D}(\omega') - \hat{D}(\omega') \hat{\rho}_{s,I}(t) \hat{D}^{\dagger}(\omega) \right) G(\omega) + h.c. \\
\end{split}
\end{equation}
where 
\begin{equation}
\label{eqn::f-and-g}
\begin{split}
F(\omega) &\equiv \frac{1}{\epsilon} \int_{0}^{\infty} d\tau\ e^{-i\omega \tau} F(t, t-\tau) 
=  \sum_{k} V_k^2 f(\eeng_k) \int_{0}^{\infty} d\tau\  e^{i(\eeng_k - \epsilon \omega) \tau}\\
 G(\omega) &\equiv \frac{1}{\epsilon} \int_{0}^{\infty} d\tau\ e^{-i\omega \tau} G(t, t-\tau) 
 = \sum_{k} V_k^2 (1-f(\eeng_k)) \int_{0}^{\infty} d\tau\  e^{i(\eeng_k - \epsilon \omega) \tau}\\
\end{split}
\end{equation}
These two equations can be viewed as the Laplace transform of time correlation functions with frequency parameter $i\omega$.

\subsection{Secular approximation}

From Equation \eqref{eqn::rf_opd}
and \eqref{eqn::f-and-g}, we observe that
$\frac{d}{dt} \hat{\rho}_{s,I}(t) =
\mathcal{O}\left(\frac{\alpha^2}{\epsilon}\right)$.
This motivates the choice of the relaxation time as
$\tau_{R} \equiv \frac{\epsilon}{\alpha^2}$.  Recall that we have
assumed $\alpha\ll \epsilon$
and $\alpha \ll 1$, then
$\alpha^2\ll \epsilon$, or equivalently
$\tau_{R} = \frac{\epsilon}{\alpha^2} \gg 1$.  Take the integral of
$\hat{\rho}_{s,I}(t)$ over time period $[t, t + r\tau_{R}]$ for $r=\mathcal{O}(1)$,
we obtain 
\[\hat{\rho}_{s,I}(t + r\tau_{R}) - \hat{\rho}_{s,I}(t) = - \frac{\alpha^2}{\epsilon} \sum_{\omega,\ \omega'} \int_{t}^{t + r\tau_{R}} ds\ e^{-i(\omega' - \omega) s} Op(s) + h.c.\]
We use the short-hand $Op(s)$ for simplicity to denote the long term
involving operators in Equation \eqref{eqn::rf_opd}. Then change the variable $s = t + \tau_{R} s'$,
\[\hat{\rho}_{s,I}(t + r\tau_{R}) - \hat{\rho}_{s,I}(t) = -  \sum_{\omega,\ \omega'} e^{-i(\omega' - \omega)t} \int_{0}^{r} ds'\  e^{-i(\omega' - \omega)\tau_{R} s'} Op(t + \tau_{R} s') + h.c.\]
By Riemann-Lebesgue lemma, if $\omega' \neq \omega$, $(\omega' - \omega) \tau_{R} = \mathcal{O}(\tau_{R}) \gg 1$, 
\[\int_{0}^{r} ds'\  e^{-i(\omega' - \omega)\tau_{R} s'} Op(t + \tau_{R} s') \approx 0. \]
Then on the right hand side, terms involving $\omega' - \omega \neq 0$ have negligible integral value. Hence,
\[\hat{\rho}_{s,I}(t + r\tau_{R}) - \hat{\rho}_{s,I}(t) \approx -  \sum_{\omega = \omega'} e^{-i(\omega' - \omega)t} \int_{0}^{r} ds'\  e^{-i(\omega' - \omega)\tau_{R} s'} Op(t + \tau_{R} s') + h.c.\]
i.e.,
\[\hat{\rho}_{s,I}(t + r\tau_{R}) - \hat{\rho}_{s,I}(t) \approx - \frac{\alpha^2}{\epsilon} \sum_{\omega = \omega'} \int_{t}^{t + r\tau_{R}} ds\ e^{-i(\omega' - \omega) s} Op(s) + h.c.\]
This is known as \emph{secular approximation} \cite{breuer}, which we have justified here in
the sense of coarse-grained approximation over relaxation time.
Divide both side by $r\tau_{R}$ and then take the limit
$r\rightarrow 0$, by fundamental theorem of calculus,
\begin{equation}
\label{eqn::ah_lindblad_sec0}
\begin{split}
\frac{d}{dt}\hat{\rho}_{s,I}(t) \approx - \frac{\alpha^2}{\epsilon} \sum_{\omega  =  \omega'}\ 
& e^{-i(\omega'-\omega) t}  \left(\hat{D}(\omega') \hat{D}^{\dagger}(\omega) \hat{\rho}_{s,I}(t) - \hat{D}^{\dagger}(\omega) \hat{\rho}_{s,I} (t) \hat{D}(\omega') \right)F(\omega) \\
+ & e^{-i(\omega'-\omega) t} \left(\hat{\rho}_{s,I}(t) \hat{D}^{\dagger}(\omega) \hat{D}(\omega') - \hat{D}(\omega') \hat{\rho}_{s,I}(t) \hat{D}^{\dagger}(\omega) \right) G(\omega) + h.c. \\
\end{split}
\end{equation}
Dropping the approximation, we arrived at the secular approximation, which is the basis for Lindblad equation:
\begin{equation}
\label{eqn::ah_lindblad_sec}
\begin{split}
\frac{d}{dt}\hat{\rho}_{s,I}(t) = -\frac{\alpha^2}{\epsilon} \sum_{\omega}  & \left(\hat{D}(\omega) \hat{D}^{\dagger}(\omega) \hat{\rho}_{s,I}(t) - \hat{D}^{\dagger}(\omega) \hat{\rho}_{s,I} (t) \hat{D}(\omega) \right)F(\omega) + \\
+ & \left(\hat{\rho}_{s,I}(t) \hat{D}^{\dagger}(\omega) \hat{D}(\omega) - \hat{D}(\omega) \hat{\rho}_{s,I}(t) \hat{D}^{\dagger}(\omega) \right) G(\omega) + h.c. \\
\end{split}
\end{equation}

\begin{remark}
By checking the previous argument, in fact, secular approximation is valid when $\frac{\alpha^2}{\eps} \ll 1$, that is,  $\alpha^2 \ll \eps$,
which appears to be a weaker condition than $\alpha \ll \eps$ used for Born-Markov approximation. 
\end{remark}

\subsection{Lindblad equation in interaction picture}

To write Equation
\eqref{eqn::ah_lindblad_sec} in a \lbd{} form, we need to decompose
coefficients $F(\omega)$ and $G(\omega)$ into their real and imaginary
parts. Let $a_{F}(\omega) := F(\omega) + F(\omega)^*$ and
$b_{F}(\omega) := \frac{F(\omega) - F(\omega)^*}{2i}$, then
$F(\omega) \equiv \frac{a_{F}(\omega)}{2} + i
b_{F}(\omega)$.
Similarly, we can decompose
$G(\omega) = \frac{\alpha_{G}(\omega)}{2} + i b_{G}(\omega)$. With
these notations, Equation \eqref{eqn::ah_lindblad_sec} becomes
\emph{\lb{} equation}
\begin{equation}
\label{eqn::ah_lindblad}
\frac{d}{dt}\hat{\rho}_{s,I} (t) = 
-\frac{i \alpha^2}{\epsilon}\bigl[\lhm_{I}, \hat{\rho}_{s,I}(t)\bigl] 
+ \frac{\alpha^2}{\epsilon} \mathcal{D}(\hat{\rho}_{s,I} (t))
\end{equation}
where \emph{\lbd\ correction Hamiltonian} $\lhm_{I}$ has the form
\begin{equation}
\lhm_{I} =  \sum_{\omega} b_{F}(\omega) \hat{D}(\omega)\hat{D}^{\dagger}(\omega) - b_{G}(\omega) \hat{D}^{\dagger}(\omega) \hat{D}(\omega)  
\end{equation}
and dissipative operator $\mathcal{D}$ has the form
\begin{equation}
\begin{split}
\mathcal{D}(\hat{\rho}_{s,I}(t)) = \sum_{\omega} 
& a_{F}(\omega) \left(\hat{D}^{\dagger}(\omega) \hat{\rho}_{s,I}(t) \hat{D}(\omega) - \frac{1}{2} \Anticom{\hat{D}(\omega) \hat{D}^{\dagger}(\omega)}{ \hat{\rho}_{s,I}(t)}\right) \\
+ & a_{G}(\omega) \left(\hat{D}(\omega) \hat{\rho}_{s,I}(t) \hat{D}^{\dagger}(\omega) - \frac{1}{2} \Anticom{\hat{D}^{\dagger}(\omega) \hat{D}(\omega) }{\hat{\rho}_{s,I}(t) }\right)
\end{split}
\end{equation}

Recall that the general dissipative operator in \lb\ equation is a linear combination of
\[
\gamma \left(\hat{L} \hat{\rho} \hat{L}^{\dagger} - \frac{1}{2}  \Anticom{\hat{L}^{\dagger} \hat{L}}{ \hat{\rho} } \right)
\]
where $\gamma$ is a constant \cite{lindblad1976}. In \ah{} model, when $\gamma = a_{F}(\omega)$ the corresponding $\hat{L} = \hat{D}^{\dagger}(\omega)$; when $\gamma = a_{G}(\omega)$, the corresponding $\hat{L} = \hat{D}(\omega)$.

\subsection{\lb{} equation in Schr{\"o}dinger picture} 
\label{subsec::lb_sch}
Transforming back into Schr{\"o}dinger picture by $\hat{\rho}_{s}(t) = e^{-\frac{i }{\epsilon}\hat{H}_s t} \hat{\rho}_{s,I}(t) e^{\frac{i }{\epsilon}\hat{H}_s t}$, we obtain \emph{\lb{} equation} in \sch{} picture,
\begin{equation}
\label{eqn::lindblad_sch}
\frac{d}{dt}\hat{\rho}_s(t) = -\frac{i}{\epsilon} \bigl[\hat{H}_s + \alpha^2 \lhm, \hat{\rho}_s(t)\bigl] + \frac{\alpha^2}{\epsilon}\mathcal{D}(\hat{\rho}_s(t))
\end{equation}
where 
\begin{equation}
\label{eqn::lindblad_hl}
\begin{split}
\lhm &= e^{-\frac{i }{\epsilon}\hat{H}_s t} \lhm_{I} e^{\frac{i }{\epsilon}\hat{H}_s t}= \lhm_{I}  = \sum_{\omega} b_{F}(\omega) \hat{D}(\omega)\hat{D}^{\dagger}(\omega) - 
b_{G}(\omega) \hat{D}^{\dagger}(\omega) \hat{D}(\omega)
\end{split}
\end{equation}
and 
\begin{equation}
\label{eqn::lindblad_diss}
\begin{split}
\mathcal{D}(\hat{\rho}_{s}(t)) = \sum_{\omega} 
& a_{F}(\omega) \left(\hat{D}^{\dagger}(\omega) \hat{\rho}_{s}(t) \hat{D}(\omega) - \frac{1}{2}  \Anticom{\hat{D}(\omega) \hat{D}^{\dagger}(\omega)}{\hat{\rho}_{s}(t) } \right) \\
+ & a_{G}(\omega) \left(\hat{D}(\omega) \hat{\rho}_{s}(t) \hat{D}^{\dagger}(\omega) - \frac{1}{2} \Anticom{\hat{D}^{\dagger}(\omega) \hat{D}(\omega)}{\hat{\rho}_{s}(t)}\right)
\end{split}
\end{equation}
Note that $\lhm$ is invariant in different pictures since $\hat{D}(\omega)$ and $\hat{D}^{\dagger}(\omega)$ both appear in the same term and thus the factor $e^{\pm i\omega t}$ will always cancel during picture transformation; this cancellation also applies to the dissipative operator $\mathcal{D}$.

To understand the Lindbladian corrected Hamiltonian, we note that
after some simple computation, it could be shown that
\[ \lhm  = \sum_{k} \left(\sum_{\omega} b_{F}(\omega) \left\vert\qinner{\phi_k^0}{\phi_{k+\omega}^1} \right\vert^2 \right) \ket{\phi_k^0,0}\bra{\phi_k^0,0} -\sum_{k} \left(\sum_{\omega} b_{G}(\omega) \left\vert\qinner{\phi_k^1}{\phi_{k-\omega}^0}\right\vert^2 \right)\ket{\phi_k^1,1}\bra{\phi_k^1,1} \]
Hence, for the new Hamiltonian, i.e.,  $\hat{H}_s + \alpha^2 \lhm$,
the set of eigenstates are the same, $\ket{\phi_k^0,0}$ and $\ket{\phi_k^1,1}$ for $k\in \Natural$. In the new Hamiltonian, the energy eigenvalues, however, are perturbed by order $\mathcal{O}(\alpha^2)$. 

Physically, this perturbation of energy comes from the interaction
between two quantum states $\ket{0}$ and $\ket{1}$ through the
coupling with bath. More specifically, since the quantum eigenstate
$\ket{\phi_k^0,0}$ interacts with quantum eigenstates
$\ket{\phi_{k+\omega}^1,1}$ (for all possible $\omega\in \Int$), their
interaction contributes to the change of energy; that is why there is
term $\left\vert\qinner{\phi_k^0}{\phi_{k+\omega}^1} \right\vert^2$.
The interaction is realized through the bath, hence the perturbed
Hamiltonian should be weighted by time correlation functions, namely,
terms $b_{F(G)}(\omega)$ and is also proportional to $\alpha^2$, the
square of the coupling parameter.

The effect of Lindblad operator will be further investigated in
Section \ref{sec::time-dependent} below in the context of perturbation
theory and it will be shown that \lb{} operator characterizes the
hopping between quantum states $\ket{0}$ and $\ket{1}$.  More specifically, the
hopping rate out of eigenstate $\ket{\phi_k^0,0}$ is
$\frac{\alpha^2}{\eps} \sum_{\omega} a_{F}(\omega)
\left\vert\qInner{\phi_k^0}{\phi_{k+\omega}^1} \right\vert^2 $.
It is worth pointing out that this expression is quite similar to the
perturbed energy eigenvalue as above,
$\alpha^2 \sum_{\omega} b_{F}(\omega)
\left\vert\qinner{\phi_k^0}{\phi_{k+\omega}^1} \right\vert^2$.
For the Laplace transform of time correlation functions, namely,
$F(\omega)$ and $G(\omega)$ in Equation \eqref{eqn::f-and-g}, the real
part contributes to (weak) hopping and the imaginary part contributes
to (weak) perturbation to the energy eigenvalues.

\subsection{Coefficients and wide band approximation}
\label{subsect::coefs}
It remains to determine the coefficients 
$a_{F,G}(\omega)$ and $b_{F,G}(\omega)$
in the \lb{} equation. Using oscillatory integral
\[\int_{0}^{\infty}d\tau \ e^{i\omega \tau} = \pi \delta(\omega) + i \pv\left(\frac{1}{\omega}\right)\]
we could obtain that
\[\begin{split}
F(\omega) &=  \sum_{k} V_k^2 f(\eeng_k) \left(  \pi \delta(\eeng_k - \epsilon \omega) + i \pv \bigl(\frac{1}{\eeng_k - \epsilon \omega}\bigl)\right)\\
G(\omega) &=   \sum_{k} V_k^2 (1-f(\eeng_k)) \left(  \pi \delta(\eeng_k - \epsilon \omega) + i \pv \bigl(\frac{1}{\eeng_k - \epsilon \omega}\bigl)\right)\\
\end{split} \]
Thus  matching the definition of $a_{F,G}(\omega)$ and $b_{F,G}(\omega)$'s
\begin{equation}\label{eqn::coefs}
\begin{split}
a_{F}(\omega) &= 2\pi \sum_{k} V_k^2 f(\eeng_k)  \delta(\eeng_k - \epsilon \omega)\\
b_{F}(\omega) &= \sum_{k} V_k^2 f(\eeng_k) \pv \bigl(\frac{1}{\eeng_k - \epsilon \omega}\bigl)\\
a_{G}(\omega) &= 2\pi  \sum_{k} V_k^2 (1-f(\eeng_k)) \delta(\eeng_k - \epsilon \omega)\\
b_{G}(\omega) &= \sum_{k} V_k^2 (1-f(\eeng_k)) \pv \bigl(\frac{1}{\eeng_k - \epsilon \omega}\bigl)\\
\end{split}\end{equation}
Notice that $a_{F,G}(\omega)$ and $b_{F,G}(\omega)$ are (generalized) functions with respect to $\omega$. Even though we only need values at $\omega\in \Int$, but these functions are indeed well-defined on $\Real$. 

In \ah{} model, we have assumed that the electron bath is infinitely
large, so continuum approximation appears to be a possible approach to
simplify the coefficients. Assume that $V^2_k \equiv V^2(\eeng_k)$ is
a continuous function of $\eeng_k$. In the discrete case, suppose the
total number of states in the bath is $N$, then $V^2(\eeng)$ should be
inversely proportional to $N$, to make the overall interaction
strength between the system and bath remain at $\Or(1)$: let
$V^2(\eeng) = \frac{\check{V}^2(\eeng)}{N}$. Let $D$ be the energy
band width of electron bath and $\nu(\eeng)$ be the density of states
at energy level $\eeng$.  In wide band approximation, to simplify
  the last equation \eqref{eqn::coefs}, it is assumed that the
  contribution to interaction strength from different energy levels of
  electron bath is approximately the same; explicitly, assume
\[2\pi \check{V}^2(\eeng) \nu(\eeng) = \Gamma\qquad \forall \eeng\in [-D,D]\]
where $\Gamma$ is a constant \cite{Bruch2016}. Then for a test function
$g(\omega)$,
\[
\begin{split}
\int_{-\infty}^{\infty}d\omega\  a_{F}(\omega) g(\omega) &= 2\pi \sum_{k} \frac{V_k^2 f(\eeng_k)}{\epsilon} g\left(\frac{\eeng_k}{\epsilon}\right)\\
&\approx 2\pi \int_{-D}^{D} d\eeng\ \check{V}^2(\eeng) \nu(\eeng) f(\eeng) \frac{1}{\epsilon} g\left(\frac{\eeng}{\epsilon}\right) \qquad\text{(Continuum approximation)}\\
&= \Gamma \int_{-D/\epsilon}^{D/\epsilon} d\omega\ f(\epsilon \omega) g(\omega) \\
&= \int_{-\infty}^{\infty}d\omega\ \Gamma \chi_{[-D/\epsilon, D/\epsilon]}(\omega) f(\epsilon \omega) g(\omega)\\
\end{split}
\]
Therefore, in the continuum limit, 
\[a_{F}(\omega) = \Gamma \chi_{[-D/\epsilon, D/\epsilon]}(\omega) f(\epsilon \omega)\]
There are two ways to get rid of the characteristic function: the first way is to  assume that $D = \infty$, mentioned in  \cite{Wenjie15_leaking}; the second way is to consider $\epsilon \downarrow 0$.  In either way, we end up with the approximation
\[a_{F}(\omega) \approx \Gamma f(\epsilon \omega)\]
These two conditions are consistent with where the term $a_{F}(\omega)$ comes. $a_{F}(\omega)$ is part of interaction strength, which involves both electron bath and open quantum system; when the elctron bath is infinitely wide or the open quantum system falls into the semi-classical region, the (generalized) function $a_{F}(\omega) $ can be approximated in this way.
Similarly, we can approximate
\[a_{G}(\omega) 
= \Gamma \chi_{[-D/\epsilon, D/\epsilon]}(\omega) (1-f(\epsilon \omega)) 
 \approx \Gamma (1-f(\epsilon \omega))
\]
For $b_{F}(\omega)$ and $b_{G}(\omega)$, we have not found easy expression for them.

\section{Semi-classical Limit}
\label{sec::semi-classical}

The semi-classical limit of \rf{} equation has been proposed and
studied in paper \cite{Wenjie15_friction, Elste2008}; the system of phase space
functions obtained in the semiclassical limit of \rf{} equation by
applying Wigner transformation is called \emph{\cme}. In the first
part, we attempt to justify the formal derivation of \cite{Wenjie15_friction} in a more
mathematical way.  In the second part, more importantly, we attempt to
study the phase space counterparts of \lb{} equation by applying
Wigner transformation.  We call the system of phase space functions 
\emph{\lcme}.  As far as we know, the Lindblad equation and its
semi-classical limit for \ah{} model have not been studied.

\subsection{Wigner transform, phase space functions and some notations}
\label{subsec::wigner_transform}
Recall that we have identified
$\epsilon \equiv \frac{\hbar\sysfreq}{\eengsc}$ as semi-classical parameter.  
Also recall that the Wigner
transformation of an operator $\hat{A}$ on $L^2(\Real)$ is defined by
\cite{wigner1932, Zworski} 
\[(\hat{A}_{\mathcal{W}})(x,p) := \int_{\Real} \hat{A} \left(x+\frac{y}{2}, x-\frac{y}{2}\right) \exp\left(-\frac{i p y  }{\eps}\right)\ dy\]
The subscript $\mathcal{W}$ indicates
the Wigner transform.
If $\hat{A} = \hat{\rho}$ is a density operator, it could be easily shown that $\int dx dp\ \frac{1}{2\pi \eps }\hat{\rho}_{\mathcal{W}}(x,p) = 1$. 
Hence if we are applying Wigner transformation to density operators $\hat{\rho}$, 
the factor $\frac{1}{2\pi \eps}$ is needed 
to have integral equal to $1$ for function $\hat{\rho}_{\mathcal{W}}$.

For a single level quantum system, 
the phase space function for a quantum master equation is clear 
(see, for instance, \cite{wigner1932}). 
As for multi-level open quantum system, the definition for phase space functions 
is not very straightforward and
we need to clarify this concept used below. 
For a general two-level open quantum system, 
the reduced density matrix can be written in the matrix form as 
\[\hat{\rho}_s(t) = \begin{bmatrix}
\hat{\rho}_{0,0}(t) & \hat{\rho}_{0,1}(t) \\
\hat{\rho}_{1,0}(t) & \hat{\rho}_{1,1}(t) \\
\end{bmatrix}\]
For a general quantum master equation, it is expected that $\hat{\rho}_{0,1}(t)$ and $\hat{\rho}_{1,0}(t)$ do not vanish.
However, it could be verified that if at time $t_0$, $\hat{\rho}_{s}(t)$ only have diagonal terms, i.e., $\hat{\rho}_{0,1}(t_0) = \hat{\rho}_{1,0}(t_0) = 0$,
then $\frac{d}{dt}\hat{\rho}_{0,1}(t_0) = \frac{d}{dt}\hat{\rho}_{1,0}(t_0) = 0$ for both \rf\ equation and \lb\ equation. Thus, $\hat{\rho}_{0,1}(t) = \hat{\rho}_{1,0}(t) = 0$ for all $t$.

Therefore, in the below, we shall only consider diagonal elements of $\hat{\rho}_s(t)$, i.e.,  assume $\hat{\rho}_s(t) = \hat{\rho}_0(t) \ket{0}\bra{0} + 
\hat{\rho}_1(t) \ket{1}\bra{1}$ for all $t$. In matrix form, 
\[\hat{\rho}_s(t) = \begin{bmatrix}
\hat{\rho}_0(t) & 0 \\
0 & \hat{\rho}_1(t) \\
\end{bmatrix}\]
The phase space function by applying Wigner transformation for
$\hat{\rho}_m(t)$ is denoted by $\wig_m(x,p,t)$, for $m \in \{0,1\}$,
namely, 
\[\wig_m(x,p,t) = \frac{1}{2\pi \eps} (\hat{\rho}_{m}(t))_{\mathcal{W}}\]
Hence, it is not difficult to show that $\int dxdp\ \wig_0(x,p,t) + \wig_1(x,p,t) = 1$.

We need to use the following lemma below.
\begin{lemma} (Semi-classical expansion, \cite[p.~66 - 68]{Zworski})
\label{lemma::semi-expan}
\begin{enumerate}
\item
  $(\hat{A}\hat{B})_{\mathcal{W}} = \hat{A}_{\mathcal{W}}
  \hat{B}_{\mathcal{W}} + \frac{i\eps}{2}
  \bigl\{\hat{A}_{\mathcal{W}}, \hat{B}_{\mathcal{W}} \bigl\} +
  \mathcal{O}(\eps^2)$
  where $\hat{A}$ and $\hat{B}$ are two
  operators. $(\hat{A}\hat{B})_{\mathcal{W}}$ is linear with respect
  to both $\hat{A}_{\mathcal{W}}$ and $\hat{B}_{\mathcal{W}}$.

\item In particular,
  $\bigl[ \hat{A}, \hat{B} \bigl]_{\mathcal{W}} = i \eps \bigl\{
  \hat{A}_{\mathcal{W}}, \hat{B}_{\mathcal{W}} \bigl\} +
  \mathcal{O}(\eps^3)$
  where $\bigl\{\cdot , \cdot \bigl\}$ is the Poisson bracket, defined
  by\footnote{We choose to follow the usual convention here, which
    differs by a negative sign compared to \cite{Zworski}}
\[\left\{h(x,p), g(x,p)\right\}  = \partial_{x} h(x,p) \partial_p g(x,p) - 
\partial_p h(x,p) \partial_x g(x,p).\]
\end{enumerate}
\end{lemma}

\subsection{Classical master equation}

As discussed in last Subsection, we shall only consider the reduced density matrix of the form $\hat{\rho}_{s}(t) =\hat{\rho}_0(t)\ket{0}\bra{0} + \hat{\rho}_1(t) \ket{1} \bra{1} $.
From Equation \eqref{eqn::redfield_sch}, we could derive that the time evolution equations for $\hat{\rho}_0(t)$ and $\hat{\rho}_1(t)$ is 
\begin{equation}
\label{eqn::redfield_sch_0}
\begin{split}
\frac{d}{dt}\hat{\rho}_0(t) =  & -\frac{i}{\epsilon} \bigl[\hat{H}_0, \hat{\rho}_0(t)\bigl] - \frac{\alpha^2}{\epsilon} \sum_{k}  V_k^2\int_{0}^{\infty} d\tau\  
e^{-i \hat{H}_1\tau}  e^{i\hat{H}_0 \tau} \hat{\rho}_0(t) e^{i \eeng_k \tau} f(\eeng_k) \\
&- \hat{\rho}_1(t) e^{-i\hat{H}_1 \tau} e^{i \hat{H}_0 \tau} e^{i \eeng_k \tau} (1-f(\eeng_k)) + h.c.\\
\end{split}
\end{equation}
and
\begin{equation}
\label{eqn::redfield_sch_1}
\begin{split}
\frac{d}{dt}\hat{\rho}_1(t) = & -\frac{i}{\epsilon}\bigl[\hat{H}_1, \hat{\rho}_1(t)\bigl] - \frac{\alpha^2}{\epsilon} \sum_{k}  V_k^2\int_{0}^{\infty} d\tau\ 
\hat{\rho}_1(t) e^{-i \hat{H}_1 \tau} e^{i \hat{H}_0 \tau} e^{i \eeng_k \tau} (1-f(\eeng_k)) \\
&- e^{-i \hat{H}_1 \tau} e^{i \hat{H}_0 \tau} \hat{\rho}_0(t) e^{i \eeng_k\tau} f(\eeng_k) + h.c.\\
\end{split}
\end{equation}
which agree with Equation (14) and (15) in Ref~\cite{Wenjie15_broadening}.

Using Lemma \ref{lemma::semi-expan}, 
we can calculate the equations 
for the corresponding Wigner transformation as
\[
\begin{split}
\partial_{t} \wig_0(t) =  \bigl\{ H_0, \wig_0(t) \bigl\} - \frac{\alpha^2}{\eps} \sum_{k} V_k^2 \int_{0}^{\infty} d\tau\ & (e^{-i \hat{H}_1 \tau})_{\mathcal{W}} (e^{i \hat{H}_0 \tau})_{\mathcal{W}} \wig_0(t) e^{i \eeng_k \tau} f(\eeng_k) \\
&  -\wig_1(t) (e^{-i \hat{H}_1 \tau})_{\mathcal{W}} (e^{i \hat{H}_0 \tau})_{\mathcal{W}}  e^{i \eeng_k \tau} (1-f(\eeng_k)) + c.c. + \mathcal{O}\left(\alpha^2 \right)\\
\partial_{t} \wig_1(t) =   \bigl\{  H_1, \wig_1(t) \bigl\} -  \frac{\alpha^2}{\epsilon} \sum_{k}  V_k^2\int_{0}^{\infty} d\tau\ & \wig_1(t)  (e^{-i \hat{H}_1 \tau})_{\mathcal{W}} (e^{i \hat{H}_0 \tau})_{\mathcal{W}} 
 e^{i \eeng_k \tau} (1-f(\eeng_k)) \\
&- (e^{-i \hat{H}_1 \tau})_{\mathcal{W}} (e^{i \hat{H}_0 \tau})_{\mathcal{W}} 
 \wig_0(t) e^{i \eeng_k\tau} f(\eeng_k) + c.c. + \mathcal{O}(\alpha^2)\\
\end{split}
\]
For clarity in equation, 
the coordinates $(x,p)$ are omitted in phase space functions 
and in $H_0$, $H_1$ as well.
After dropping higher order terms of $\mathcal{O}(\alpha^2)$ 
\begin{equation}
\begin{split}
\partial_{t} \wig_0 &=  \bigl\{ H_0, \wig_0 \bigl\} -  \gamma_{0\rightarrow 1} \wig_0 + \gamma_{1\rightarrow 0} \wig_1 \\
\partial_{t} \wig_1 &=   \bigl\{  H_1, \wig_1 \bigl\} + \gamma_{0\rightarrow 1} \wig_0 - \gamma_{1\rightarrow 0} \wig_1 \\
\end{split}
\end{equation}
where hopping rates 
\begin{equation}
\begin{split}
\gamma_{0\rightarrow 1} &= \frac{\alpha^2}{\eps} \sum_{k} V_k^2 \int_{0}^{\infty} d\tau\ (e^{-i \hat{H}_1 \tau})_{\mathcal{W}} (e^{i \hat{H}_0 \tau})_{\mathcal{W}} e^{i \eeng_k \tau} f(\eeng_k) + c.c. \\
\gamma_{1\rightarrow 0} &= \frac{\alpha^2}{\eps} \sum_{k} V_k^2 \int_{0}^{\infty} d\tau\ (e^{-i \hat{H}_1 \tau})_{\mathcal{W}} (e^{i \hat{H}_0 \tau})_{\mathcal{W}} e^{i \eeng_k \tau} (1- f(\eeng_k) )+ c.c.\\
\end{split}
\end{equation}
Notice that hopping rates are functions of phase space coordinates $x$ and $p$. They describe how fast the  jumping between states $\ket{0}$ and $\ket{1}$ depending on $(x, p)$.

\begin{remark}
  Compared with the result in \cite{Wenjie15_friction}, the rates
  $\gamma_{0\to1}$ and $\gamma_{1 \to 0}$ we have above are
  considerably more complicated. Here we provide a heuristic
  simplification of the expression, though we do not know how to
  justify the argument on a more rigorous level.

  Equation (60) in \cite{curtright} shows that if $\hat{H}$ is a
  Hamiltonian for harmonic oscillators,
\[(e^{i\hat{H}t})_{\mathcal{W}} = \cos(t/2)^{-1} e^{2i H \tan(t/2)}\]
where $H = (\hat{H})_{\mathcal{W}}$.
Since $\hat{H}_0$ and $\hat{H}_1$ are Hamiltonian for harmonic oscillators,
by using the last equation,
\[
\begin{split}
(e^{-i \hat{H}_1 \tau})_{\mathcal{W}} (e^{i \hat{H}_0 \tau})_{\mathcal{W}}  &= \cos(\tau/2)^{-2} e^{-2i (H_1 - H_0) \tan(\tau/2)} \\
&= \cos(\tau/2)^{-2} e^{-2i \epe(x) \tan(\tau/2)} \\
& \approx e^{-i \epe(x) \tau} \qquad \text{when }  \tau \downarrow 0
\end{split}
\]
In the last step, we use $\cos(\tau/2)\rightarrow 1$ and $\tan(\tau/2)\rightarrow \tau/2$ as $\tau\rightarrow 0$.
Then, hopping rates can be written as
\[\gamma_{0\rightarrow 1} = \frac{\alpha^2}{\eps} \sum_{k} V_k^2 \int_{0}^{\infty} d\tau\  \cos(\tau/2)^{-2} e^{-2i \epe(x) \tan(\tau/2)} e^{i \eeng_k \tau} f(\eeng_k) + c.c.
\]
Heuristically, if we approximate $(e^{-i \hat{H}_1 \tau})_{\mathcal{W}} (e^{i \hat{H}_0 \tau})_{\mathcal{W}}$ by $e^{-i \epe (x) \tau}$, i.e., assume that the integral is mostly contributed from $\tau$ near $0$, then 
\[\begin{split}
\gamma_{0\rightarrow 1} &= \frac{\alpha^2}{\eps} 2\pi \sum_{k} V_k^2 f(\eeng_k)  \delta(\eeng_k - \epe(x))\\
&=  \frac{\alpha^2}{\eps}\Gamma f(\epe(x)) \chi_{[\epe^{-1}(-D), \epe^{-1}(D)]}(x)\\
&\approx \frac{\alpha^2}{\eps} \Gamma f(\epe(x))\qquad (\text{let } D=\infty, \text{ i.e., wide band approximation})
\end{split}\]
which becomes the result in \cite{Wenjie15_friction}.  $U^{-1}(x)$
is the inverse function of $U(x)$; since $U(x)$ is a linear function,
$U^{-1}(x)$ is well-defined.  The second and third step of last
equation use similar computation as wide band approximation, which has
been shown in details in Section \ref{subsect::coefs}. This heuristic
computation gives a nice simple expression for the rates, but it
should be pointed out we do not know how to justify the crucial
approximation above of $(e^{-i \hat{H}_1 \tau})_{\mathcal{W}} (e^{i \hat{H}_0 \tau})_{\mathcal{W}}$ by $e^{-i \epe (x) \tau}$. 
\end{remark}

\subsection{\lcme}

Recall that if we assume at time $t_0$, $\hat{\rho}_{s}(t_0)$ is diagonal, 
then reduced density operator $\hat{\rho}_{s}(t) = \hat{\rho}_0(t)\ket{0}\bra{0} + \hat{\rho}_1(t)\ket{1}\bra{1}$
without terms involving $\ket{1}\bra{0}$ nor $\ket{0}\bra{1}$.
The time-evolution \lb{} equations can be written more explicitly as
\begin{equation}
\label{eqn::lindblad_sch_diag_0}
\begin{split}
\frac{d}{dt}\hat{\rho}_0(t) = & -\frac{i}{\epsilon} \bigl[\hat{H}_0 + \alpha^2 \sum_{\omega} b_{F}(\omega) \bra{0} \hat{D}(\omega) \ket{1} \bra{1} \hat{D}^{\dagger}(\omega) \ket{0}, \hat{\rho}_0(t) \bigl] \\
& + \frac{\alpha^2}{\epsilon}\sum_{\omega} a_{G}(\omega) \bra{0} \hat{D}(\omega) \ket{1} \hat{\rho}_1(t) \bra{1} \hat{D}^{\dagger}(\omega) \ket{0} \\
& - \frac{\alpha^2}{\epsilon}\sum_{\omega} \frac{a_{F}(\omega)}{2}  \left( \bra{0} \hat{D}(\omega) \ket{1} \bra{1} \hat{D}^{\dagger}(\omega) \ket{0} \hat{\rho}_0(t) +  \hat{\rho}_0(t) \bra{0} \hat{D}(\omega) \ket{1} \bra{1} \hat{D}^{\dagger}(\omega) \ket{0}\right)
\end{split}
\end{equation}

\begin{equation}
\label{eqn::lindblad_sch_diag_1}
\begin{split}
\frac{d}{dt}\hat{\rho}_1(t) = & -\frac{i}{\epsilon}\bigl[  \hat{H}_1 -\alpha^2 \sum_{\omega} b_{G}(\omega) \bra{1} \hat{D}^{\dagger}(\omega) \ket{0} \bra{0} \hat{D}(\omega) \ket{1} , \hat{\rho}_1(t) \bigl] \\
& + \frac{\alpha^2}{\epsilon}\sum_{\omega} a_{F}(\omega) \bra{1} \hat{D}^{\dagger}(\omega) \ket{0} \hat{\rho}_0(t) \bra{0} \hat{D}(\omega) \ket{1} \\
& - \frac{\alpha^2}{\epsilon}\sum_{\omega} \frac{a_{G}(\omega)}{2}
\left( \bra{1} \hat{D}^{\dagger}(\omega) \ket{0} \bra{0} \hat{D}(\omega) \ket{1} \hat{\rho}_1(t) + \hat{\rho}_1(t) \bra{1} \hat{D}^{\dagger}(\omega) \ket{0} \bra{0} \hat{D}(\omega) \ket{1} \right)
\end{split}
\end{equation}

The system of time-evolution equations obtained by applying Wigner
transformation to \lb{} equation is given by, after some
straightforward calculations
\begin{equation}
\label{eqn::lcme0}
\begin{split}
\partial_t \wig_0(t) &= \bigl\{ H_0 + \alpha^2 \lhms_0, \wig_0(t) \bigl\} \\
& + \frac{\alpha^2}{\eps} \sum_{\omega} a_{G}(\omega) \left(\bra{0} \hat{D}(\omega)  \hat{D}^{\dagger}(\omega) \ket{0}\right)_{\mathcal{W}} \wig_1(t)  \\
& - \frac{\alpha^2}{\epsilon}\sum_{\omega} a_{F}(\omega)  \left(\bra{0} \hat{D}(\omega) \hat{D}^{\dagger}(\omega) \ket{0}\right)_{\mathcal{W}} \wig_0(t) +\mathcal{O}(\alpha^2, \alpha^2\eps^2) \\
\end{split}
\end{equation}
\begin{equation}
\label{eqn::lcme1}
\begin{split}
\partial_t \wig_1(t) &= \bigl\{ H_1 - \alpha^2 \lhms_1, \wig_1(t) \bigl\}  \\
& + \frac{\alpha^2}{\epsilon}\sum_{\omega} a_{F}(\omega) \left(\bra{1} \hat{D}^{\dagger}(\omega)  \hat{D}(\omega) \ket{1} \right)_{\mathcal{W}} \wig_0(t)  \\
& - \frac{\alpha^2}{\epsilon}\sum_{\omega} a_{G}(\omega)
\left( \bra{1} \hat{D}^{\dagger}(\omega) \hat{D}(\omega) \ket{1}\right)_{\mathcal{W}} \wig_1(t) + \mathcal{O}(\alpha^2, \alpha^2\eps^2)\\
\end{split}\end{equation}
where 
\[\lhms_0 = \sum_{\omega} b_{F}(\omega) \left(\bra{0} \hat{D}(\omega) \hat{D}^{\dagger}(\omega) \ket{0}\right)_{\mathcal{W}}, \quad \text{and} \quad 
\lhms_1 = \sum_{\omega} b_{G}(\omega) \left(\bra{1} \hat{D}^{\dagger}(\omega) \hat{D}(\omega) \ket{1}\right)_{\mathcal{W}}\]

The error terms with order $\mathcal{O}(\alpha^2)$ come from the Wigner transformation of hopping terms 
and the error terms with order $\mathcal{O}(\alpha^2\eps^2)$ come from Wigner transformation of commutators.

\section{Comparison of \rf\ equation and \lb\ equation from \pertt}
\label{sec::time-dependent}

In this section, we intend to use \pertt{}
to understand the similarity and difference of \rf{} equation and
\lb{} equation, by considering the hopping between different quantum
states.  The mathematical tool for \pertt{} has
been widely studied, and one of the most famous formulas from that is
Fermi Golden Rule. 

The main finding is that if the quantum system is prepared at a pure
state $\ket{\phi_i^0, 0}$, then its hopping rates to other quantum
states $\ket{\phi_f^0,0}$ and $\ket{\phi_f^1,1}$ are the same up to
the first order, for \rf{} and \lb{} equations. By linearity of \rf{}
and \lb{} equation, this conclusion is also true if initially
$\hat{\rho}_{m}$ ($m = 0, 1$) are diagonal.

\subsection{Perturbation of \rf{} and \lb\ equations}
As discussed above, \lb{} equation can be derived by using secular approximation from \rf{} equation, under the condition that $\alpha^2\ll \eps$.
Under this condition, the hopping term in \rf{} equation 
(Equation \eqref{eqn::rf_opd}) could be considered as a small perturbation, that is, $\frac{\alpha^2}{\eps}$ is regarded as perturbation parameter below.

In \sch{} picture, the \rf{} equation has the form, by using Equation \eqref{eqn::rf_opd},
\begin{equation}
\label{eqn::redfield_sch_opd}
\begin{split}
\frac{d}{dt}\hat{\rho}_{s}(t) = -\frac{i}{\eps}\bigl[\hat{H}_s, \hat{\rho}_s(t) \bigl] 
-\frac{\alpha^2}{\eps} \sum_{\omega, \omega'} 
& \left(\hat{D}(\omega') \hat{D}^{\dagger}(\omega) \hat{\rho}_{s}(t) 
- \hat{D}^{\dagger}(\omega) \hat{\rho}_{s} (t) \hat{D}(\omega') \right)F(\omega) \\
 + & \left(\hat{\rho}_{s}(t) \hat{D}^{\dagger}(\omega) \hat{D}(\omega') - \hat{D}(\omega') \hat{\rho}_{s}(t) \hat{D}^{\dagger}(\omega) \right) G(\omega) + h.c. \\
\end{split}
\end{equation}

After rewriting it in terms of $a_{F(G)}(\omega)$ and $b_{F(G)}(\omega)$, it becomes
\begin{equation}
\label{eqn::redfield_sch_opd2}
\frac{d}{dt}\hat{\rho}_{s}(t) = -\frac{i}{\eps}\bigl[\hat{H}_s, \hat{\rho}_s(t) \bigl]  + \frac{\alpha^2}{\eps}\mathcal{R}(\hat{\rho}_s(t))
\end{equation}
where operator $\mathcal{R}$ has the form 
\begin{equation}
\begin{split}
\mathcal{R}(\hat{\rho}_s(t)) = 
& -\frac{1}{2} \sum_{\omega, \omega'} a_{F}(\omega) 
\left(
\hat{D}(\omega') \hat{D}^{\dagger}(\omega) \hat{\rho}_{s}(t) 
- \hat{D}^{\dagger}(\omega) \hat{\rho}_{s} (t) \hat{D}(\omega') 
+ h.c. \right) \\
& -\frac{1}{2}\sum_{\omega, \omega'} a_{G}(\omega) 
\left(\hat{\rho}_{s}(t) \hat{D}^{\dagger}(\omega) \hat{D}(\omega') - \hat{D}(\omega') \hat{\rho}_{s}(t) \hat{D}^{\dagger}(\omega) + h.c. \right)\\
&-i \sum_{\omega, \omega'} b_{F}(\omega) \left(
\hat{D}(\omega') \hat{D}^{\dagger}(\omega) \hat{\rho}_{s}(t) 
- \hat{D}^{\dagger}(\omega) \hat{\rho}_{s} (t) \hat{D}(\omega') 
- h.c. \right)\\
&-i \sum_{\omega, \omega'} b_{G}(\omega) 
\left(\hat{\rho}_{s}(t) \hat{D}^{\dagger}(\omega) \hat{D}(\omega') - \hat{D}(\omega') \hat{\rho}_{s}(t) \hat{D}^{\dagger}(\omega) 
- h.c. \right)\\
\end{split}
\end{equation}
If we only keep those terms with $\omega' = \omega$, it becomes  \lb{} equation, that is Equation \eqref{eqn::lindblad_sch}.

Then, consider the \pert{} for the system initially at
$\ket{\phi_{i}^{0}, 0}$ and ending at $\ket{\phi_{f}^{0}, 0}$ or
$\ket{\phi_{f}^{1}, 1}$.  We are interested in finding the transition
between different quantum states. Because of the symmetry of this
problem, it suffices to look at transition from $\ket{\phi_i^{0}, 0}$
to $\ket{\phi_f^0, 0}$ and to $\ket{\phi_f^{1},1}$.

Assume that 
\[\hat{\rho}_{s}(t) = \hat{\rho}_{s,0}(t) + \frac{\alpha^2}{\eps} \hat{\rho}_{s,1}(t) + \mathcal{O}\left(\bigl(\frac{\alpha^2}{\eps}\bigl)^2\right)\]
Then by collecting terms of the same order, 
\[\begin{split}
\mathcal{O}(1) :\qquad & \frac{d}{dt} \hat{\rho}_{s,0}(t) = -\frac{i}{\eps} \bigl[\hat{H}_s, \hat{\rho}_{s,0}(t)\bigl], \qquad \hat{\rho}_{s,0}(0) = \ket{\phi_i^0, 0}\bra{\phi_i^0,0} \equiv \Pi_{i}^{(0)} \\
\mathcal{O}\left(\frac{\alpha^2}{\eps}\right): \qquad & 
\frac{d}{dt}\hat{\rho}_{s,1}(t) = -\frac{i}{\eps} \bigl[\hat{H}_s, \hat{\rho}_{s,1}(t) \bigl]
+ \frac{\alpha^2}{\eps}\mathcal{R}(\hat{\rho}_{s,0}(t)), \qquad \hat{\rho}_{s,1}(0) = 0\\
\end{split}\]

For the leading order, the solution is trivial to find $\hat{\rho}_{s,0}(t) = \ket{\phi_i^0, 0}\bra{\phi_i^0,0} \equiv \Pi_{i}^{(0)}$. For the first order, after some computing, 
\[\begin{split}
\mathcal{R}(\hat{\rho}_{s,0}(t)) = -\frac{1}{2}\sum_{\omega, \omega'} a_{F}(\omega) 
\bigl(& 
\Pi_{i+\omega-\omega'}^{(0)} \hat{d} \Pi^{(1)}_{i+\omega} \hat{d}^{\dagger} \Pi_{i}^{(0)}
- \Pi^{(1)}_{i+\omega} \hat{d}^{\dagger} \Pi_{i}^{(0)} \hat{d} \Pi_{i+\omega'}^{(1)} \\
&
+ \Pi^{(0)}_{i} \hat{d} \Pi_{i+\omega}^{(1)} \hat{d}^{\dagger} \Pi_{i+\omega-\omega'}^{(0)}
- \Pi_{i+\omega'}^{(1)} \hat{d}^{\dagger} \Pi_{i}^{(0)} \hat{d} \Pi_{i+\omega}^{(1)}
\bigl)\\
- i\sum_{\omega, \omega'} b_{F}(\omega)
\bigl(& 
\Pi_{i+\omega-\omega'}^{(0)} \hat{d} \Pi^{(1)}_{i+\omega} \hat{d}^{\dagger} \Pi_{i}^{(0)}
- \Pi^{(1)}_{i+\omega} \hat{d}^{\dagger} \Pi_{i}^{(0)} \hat{d} \Pi_{i+\omega'}^{(1)} \\
&
- \Pi^{(0)}_{i} \hat{d} \Pi_{i+\omega}^{(1)} \hat{d}^{\dagger} \Pi_{i+\omega-\omega'}^{(0)}
+ \Pi_{i+\omega'}^{(1)} \hat{d}^{\dagger} \Pi_{i}^{(0)} \hat{d} \Pi_{i+\omega}^{(1)}
\bigl)\\
\end{split} \]
The terms involving $a_{G}(\omega)$ and $b_{G}(\omega)$ vanish, since there is no hopping from state $\ket{1}$ to $\ket{0}$ in our setting up.
To study transition between different quantum states, 
let 
\begin{equation}
\lambda_{f}^{(m)}(t) \equiv \Bra{\phi_f^0, 0}\hat{\rho}_{s,m}(t) \Ket{\phi_f^0,0}\qquad \theta_{f}^{(m)}(t) \equiv \Bra{\phi_f^1, 1}\hat{\rho}_{s,m}(t) \Ket{\phi_f^1,1}\end{equation}
where $m = 0,1, 2, \cdots$ represents the perturbation order. 
For the leading order, apparently,
$\lambda_{f}^{(0)}(t) = \delta_{i,f}$ and $\theta^{(0)}_{f}(t) = 0$ for $f\in \Natural$. 

As for the first-order, it could be obtained that
\begin{equation}
\label{eqn::lambda_special}
\begin{split}
\frac{d}{dt}\lambda_{f}^{(1)}(t) &= \frac{\alpha^2}{\eps} \Bra{\phi_f^0,0} \mathcal{R}(\hat{\rho}_{s,0}(t))\Ket{\phi_f^0,0} \\
&=  \frac{\alpha^2}{\eps} \left( - \sum_{\omega, \omega'} a_{F}(\omega) \delta_{i,f} \delta_{\omega, \omega'}  \left\vert\qInner{\phi_i^0}{\phi_{i+\omega}^1} \right\vert^2 \right)\\
&= -\frac{\alpha^2}{\eps} \sum_{\omega} a_{F}(\omega) \left\vert\qInner{\phi_i^0}{\phi_{i+\omega}^1} \right\vert^2 \delta_{i,f}
\end{split}
\end{equation}
and
\begin{equation}
\label{eqn::theta_special}
\begin{split}
\frac{d}{dt}\theta_{f}^{(1)}(t) &= \frac{\alpha^2}{\eps} \Bra{\phi_f^1,1} \mathcal{R}(\hat{\rho}_{s,0}(t))\Ket{\phi_f^1,1} \\
&= \frac{\alpha^2}{\eps} \sum_{\omega, \omega'} a_{F}(\omega) \delta_{\omega, \omega'} \delta_{f, i+\omega} \left\vert\qInner{\phi_i^0}{\phi_{f}^1} \right\vert^2 \\
&= \frac{\alpha^2}{\eps}  a_{F}(f-i) \left\vert\qInner{\phi_i^0}{\phi_{f}^1} \right\vert^2
\end{split}
\end{equation}
As we observe, for the diagonal elements of
$\mathcal{R}(\hat{\rho}_{s,0}(t))$, the only contribution comes from
terms with $\omega = \omega'$.  Therefore, if this condition
$\omega' = \omega$ is imposed beforehand, the result for
$\frac{d}{dt}\lambda_{f}^{(1)}(t)$ and
$\frac{d}{dt}\theta_{f}^{(1)}(t)$ will not change.  This implies that
if we are applying perturbation to \lb\ equation, we should obtain
exactly the same set of equations for the diagonal elements up to the
first order.

It could be noticed that the rate of probability leaving $\ket{\phi_i^0,0}$ is the summation of rate of probability entering $\ket{\phi_f^1,1}$ for varying $f\in \Natural$.
The rate of leaving $\ket{\phi_i^0,0}$ is 
$\frac{\alpha^2}{\eps}\sum_{\omega} a_{F}(\omega) \left|\bran{\phi_i^0}\ket{\phi_{i+\omega}^1}\right|^2$
and it hops to state $\ket{\phi_f^1,1}$ with rate 
$\frac{\alpha^2}{\eps} a_{F}(\omega) \left| \bran{\phi_i^0} \ket{\phi_{i+\omega}^1}  \right|^2$ if $f = i+\omega$ for some $\omega\in \Int$.

\subsection{More general initial condition}

A slightly more general initial condition is to assume that $\hat{\rho}_{0}$ and $\hat{\rho}_1$ are diagonal, i.e., 
$\hat{\rho}_{0}(0) = \sum_{k} \lambda_{k} \ket{\phi_k^0}\bra{\phi_k^0}$ and $\hat{\rho}_{1}(0) = \sum_{k} \theta_{k} \ket{\phi_k^1}\bra{\phi_k^1}$.
Then the leading order is 
\[\lambda_{k}^{(0)}(t) = \lambda_k \qquad \theta_{k}^{(0)}(t) = \theta_k\qquad \forall k \in \Natural,\ t \ge 0 \]

Then we need to compute the first order,
\[\begin{split}
\mathcal{R}(\hat{\rho}_{s,0}(t)) = -\frac{1}{2}\sum_{\omega, \omega', i} \lambda_i a_{F}(\omega) 
\bigl(& 
\Pi_{i+\omega-\omega'}^{(0)} \hat{d} \Pi^{(1)}_{i+\omega} \hat{d}^{\dagger} \Pi_{i}^{(0)}
- \Pi^{(1)}_{i+\omega} \hat{d}^{\dagger} \Pi_{i}^{(0)} \hat{d} \Pi_{i+\omega'}^{(1)} + h.c. \bigl)\\
- i\sum_{\omega, \omega', i} \lambda_i b_{F}(\omega)
\bigl(& 
\Pi_{i+\omega-\omega'}^{(0)} \hat{d} \Pi^{(1)}_{i+\omega} \hat{d}^{\dagger} \Pi_{i}^{(0)}
- \Pi^{(1)}_{i+\omega} \hat{d}^{\dagger} \Pi_{i}^{(0)} \hat{d} \Pi_{i+\omega'}^{(1)} -h.c. \bigl)\\
- \frac{1}{2}\sum_{\omega, \omega', i} \theta_{i} a_{G}(\omega) \bigl(
& \Pi_{i}^{(1)} \hat{d}^{\dagger} \Pi_{i-\omega}^{(0)} \hat{d} \Pi_{i-\omega+\omega'}^{(1)}
- \Pi_{i-\omega'}^{(0)} \hat{d} \Pi_{i}^{(1)} \hat{d}^{\dagger} \Pi_{i-\omega}^{(0)} + h.c. \bigl) \\
- i \sum_{\omega, \omega', i} \theta_{i} b_{G}(\omega) \bigl(
& \Pi_{i}^{(1)} \hat{d}^{\dagger} \Pi_{i-\omega}^{(0)} \hat{d} \Pi_{i-\omega+\omega'}^{(1)}
- \Pi_{i-\omega'}^{(0)} \hat{d} \Pi_{i}^{(1)} \hat{d}^{\dagger} \Pi_{i-\omega}^{(0)} - h.c. \bigl) \\
\end{split} \]

Then, after some computation
\begin{equation}
\label{eqn::lambda}
\begin{split}
\frac{d}{dt}\lambda_{f}^{(1)}(t) &= \frac{\alpha^2}{\eps} \Bra{\phi_f^0,0} \mathcal{R}(\hat{\rho}_{s,0}(t))\Ket{\phi_f^0,0} \\
&=  \frac{\alpha^2}{\eps} \left( - \sum_{\omega, \omega', i} \lambda_i a_{F}(\omega) \delta_{i,f} \delta_{\omega, \omega'}  \left\vert\qInner{\phi_i^0}{\phi_{i+\omega}^1} \right\vert^2 + \sum_{\omega, \omega', i}\theta_i a_{G}(\omega) \delta_{i-\omega, f} \delta_{\omega, \omega'} \left\vert \qInner{\phi_f^0}{\phi_i^1} \right\vert^2 \right)\\
&= -\frac{\alpha^2}{\eps} \sum_{\omega} \lambda_f a_{F}(\omega) \left\vert\qInner{\phi_f^0}{\phi_{f+\omega}^1} \right\vert^2 + \frac{\alpha^2}{\eps}
\sum_{i} \theta_{i} a_{G}(i-f) \left\vert \qInner{\phi_f^0}{\phi_{i}^1} \right\vert^2
\end{split}
\end{equation}
and
\begin{equation}
\label{eqn::theta}
\begin{split}
\frac{d}{dt}\theta_{f}^{(1)}(t) &= \frac{\alpha^2}{\eps} \Bra{\phi_f^1,1} \mathcal{R}(\hat{\rho}_{s,0}(t))\Ket{\phi_f^1,1} \\
&= \frac{\alpha^2}{\eps}\left( \sum_{\omega, \omega', i} \lambda_i a_{F}(\omega) \delta_{\omega, \omega'} \delta_{f, i+\omega} \left\vert\qInner{\phi_i^0}{\phi_{f}^1} \right\vert^2  
- \sum_{\omega, \omega', i} \theta_{i} a_{G}(\omega)\delta_{i,f} \delta_{\omega,\omega'} \left\vert \qInner{\phi_i^1}{\phi_{i-\omega}^0}  \right\vert^2 \right)\\
&= \frac{\alpha^2}{\eps}  \sum_{i}\lambda_i a_{F}(f-i) \left\vert\qInner{\phi_i^0}{\phi_{f}^1} \right\vert^2 - \frac{\alpha^2}{\eps}\sum_{\omega} \theta_{f} a_{G}(\omega)  \left\vert \qInner{\phi_f^1}{\phi_{f-\omega}^0}  \right\vert^2
\end{split}
\end{equation}

Hence the hopping rate from $\ket{\phi_i^0, 0}$ to
$\ket{\phi_{f}^1, 1}$ is
$\frac{\alpha^2}{\eps}a_{F}(f-i) \left| \langle \phi_{i}^0 | \phi_{f}^1 \rangle \right|^2$
and hopping rate from $\ket{\phi_i^1, 1}$ to $\ket{\phi_f^0, 0}$ is
$\frac{\alpha^2}{\eps} a_{G}\left(i-f\right) \left| \langle \phi_f^0 | \phi_i^1 \rangle
\right|^2$.
Because $\rhoo_{k}(t)$ and $\theta_{k}(t)$ have an interpretation
as the probability at state $\ket{\phi_k^0, 0}$ or
$\ket{\phi_k^1, 1}$, then Equations \eqref{eqn::lambda} and \eqref{eqn::theta} can
be interpreted as the Kolmogorov's backward equation for a continuous
time Markov Chain. To make sure that the first-order perturbation is
valid,  the time horizon is assumed to be
$t \ll \frac{\eps}{\alpha^2}$.

The computation shows that if initially $\hat{\rho}_{0}(0)$ and $\hat{\rho}_{1}(0)$ are diagonal, then \rf{} and \lb{} equation lead to the same transition rate up to the first order. 
Since the operator $\mathcal{R}$ is linear with respect to the input density operator,
this result is just a natural extension of that in last Subsection.

\subsection{Interpretation of hopping rate in LCME}

We continue to assume that $\hat{\rho}_{m}(0)$ ($m=0,1$) are diagonal and in this Subsection, 
we will consider \lb\ equation only.
It could be easily verified that $\hat{\rho}_{m}(t)$ are diagonal for any time $t\ge 0$.
Hence
\[\hat{\rho}_0(t) = \sum_{k} \lambda_k(t) \ket{\phi_{k}^0}\bra{\phi_k^0}\]
Consider the jumping leaving $\ket{0}$ to $\ket{1}$ in LCME, that is, 
\[\frac{\alpha^2}{\eps} \sum_{\omega} a_{F}(\omega) \left(\bra{0} \hat{D}(\omega) \hat{D}^{\dagger}(\omega) \ket{0} \right)_{\mathcal{W}} \wig_0(x, p, t)\]
 in Equation \eqref{eqn::lcme0}.
By using definition of $\hat{D}^{(\dagger)}(\omega)$ in Equation \eqref{eqn::defn_D}, and using Lemma \ref{lemma::semi-expan}, we could obtain 
\begin{equation}
\begin{split}
& \frac{\alpha^2}{\eps} \sum_{\omega} a_{F}(\omega) \left(\bra{0} \hat{D}(\omega) \hat{D}^{\dagger}(\omega) \ket{0} \right)_{\mathcal{W}} \wig_0(x, p, t) \\
= & \frac{\alpha^2}{\eps} \sum_{\omega} a_{F}(\omega) \left(\bra{0} \hat{D}(\omega) \hat{D}^{\dagger}(\omega) \ket{0} \frac{1}{2\pi \eps} \hat{\rho}_0 \right)_{\mathcal{W}}  + \mathcal{O}(\alpha^2) \\
= &  \frac{\alpha^2}{\eps} \sum_{\omega} a_{F}(\omega) 
\sum_{k} \left| \bra{\phi_{k}^0} \phi^1_{k+\omega} \rangle  \right| ^2 
\lambda_{k}(t) \frac{1}{2\pi\eps}\left(\ket{\phi_{k}^0} \bra{\phi_{k}^0}\right)_{\mathcal{W}} + \mathcal{O}(\alpha^2)\\
=& \sum_{k}  \left(\frac{\alpha^2}{\eps} \sum_{\omega} a_{F}(\omega) \left| \bra{\phi_{k}^0} \phi^1_{k+\omega} \rangle  \right| ^2 \right) \lambda_k(t) \frac{1}{2\pi \eps}\left(\ket{\phi_{k}^0} \bra{\phi_{k}^0}\right)_{\mathcal{W}} + \mathcal{O}(\alpha^2) 
\end{split}
\end{equation}
This equation shows that the hopping rate from state $\ket{0}$ to $\ket{1}$ in semi-classical limit is the summation of the contribution from each state $\ket{\phi_{k}^0, 0}$. More specifically, 
the probability at state $\ket{\phi_k^0,0}$ in the semi-classical sense is 
$\lambda_k(t) \frac{1}{2\pi \eps}\left(\ket{\phi_{k}^0} \bra{\phi_{k}^0}\right)_{\mathcal{W}}$; the hopping rate out of state $\ket{\phi_k^0}$ is $\left(\frac{\alpha^2}{\eps} \sum_{\omega} a_{F}(\omega) \left| \bra{\phi_{k}^0} \phi^1_{k+\omega} \rangle  \right| ^2 \right)$; their product is exactly the contribution from quantum state $\ket{\phi_k^0,0}$.

Notice that the hopping rate obtained here is consistent with Equation \eqref{eqn::lambda_special} obtained by perturbation theory.
This matches our intuition and it connects the hopping rate in LCME with the hopping rate from \lb{} equation.

\subsection{Discussion on Franck-Condon Blockade}

Using the wide band approximation, the hopping rate could be explicitly computed using Franck-Condon factors \cite{Jens2005, Wenjie15_broadening}. The Franck-Condon factor is
\[\qInner{\phi_n^0}{\phi_m^1} = \frac{(-1)^{n-N} \sqrt{N!}}{\sqrt{M!}} \exp\left(-\frac{g^2}{2\eps}\right) \left(\frac{g}{\sqrt{\eps}}\right)^{M-N} L_{N}^{(M-N)}\left(\frac{g^2}{\eps}\right)\]
where $L_{k}^{(k')}(x)$ is the generalized Laguerre polynomial, $N = \min(n,m)$ and $M = \max(n,m)$. 
Then the hopping rate from $\ket{\phi_n^0,0}$ to $\ket{\phi_m^1,1}$ is
\[
\frac{\alpha^2 \Gamma}{\eps} \frac{1}{1+e^{\beta\eps(m-n)}} \frac{N!}{M!} \exp\left(-\frac{g^2}{\eps} \right) \left(\frac{g^2}{\eps}\right)^{M-N} \left( L_N^{(M-N)}\left(\frac{g^2}{\eps}\right)\right)^2
\]
We observe that the hopping rate is small due to the weak
phonon-reservoir coupling (\ie, $\frac{\alpha^2}{\eps} \ll 1$) and
also since the Franck-Condon factor $\exp(-g^2/\eps)$ becomes
exponentially small as $g^2/\eps\rightarrow \infty$, known as the
Franck-Condon blockade \cite{Jens2005, Dmitry} for large on-the-molecule
electron-phonon coupling (\ie, $g \gg 1$). From the last expression,
we notice that the Franck-Condon blockade occurs when the ratio of
electron-phonon coupling rate $g^2$ and semi-classical parameter
$\eps$ is large; in particular, even for finite $g$, if
$\eps\rightarrow 0$, such exponentially small hopping rate also
appears.

\section{Discussion and Conclusion}
\label{sec::conclusion}

In this paper, we have revisited the derivation of \rf{} equation by
using time-convolutionless equation (TCL) and consequently derive the
\lb{} equation using secular approximation, in the context of
Anderson-Holstein model.  As an analogy to \cme, \lcme\ (LCME) is
introduced and its form is given in Equations \eqref{eqn::lcme0} and
\eqref{eqn::lcme1}. The comparison between \rf{} equation and \lb{}
equation from the perspective of \pertt{} is considered: \rf{}
equation and \lb{} equation yield the same hopping rate in the first
order; in other words, the dynamics of their diagonal elements of
reduced density operator $\hat{\rho}_s$ are the same, to the first
order.
The condition of the derivation and perturbation result both suggest
that \lb{} equation might be a better candidate for studying \ah{}
model than \rf{} equation and reasons are listed as follow: first,
from the derivation, they are at the same level, i.e., both under
weak-coupling limit $\alpha \ll \eps$; second, the condition of
deriving \lb{} equation from \rf{} equation is redundant and there is,
in fact, no further constraint in approximating \rf{} equation by
\lb{} equation; third, they have the same hopping rates between
eigenstates up to the first order in perturbation theory; finally, the
analysis for \lb{} equation is easier.

Usually, for multi-level open quantum systems, the density operator 
$\hat{\rho}_s(t)$ does not necessarily have vanishing $\hat{\rho}_{0,1}(t)$ and $\hat{\rho}_{1,0}(t)$. These two off-diagonal operators will cause more challenge in both analysis and numerics. 
As we have shown, for \rf{} equation, if we start from diagonal density operator, i.e., 
$\hat{\rho}_s(t) = \hat{\rho}_0(t) \ket{0}\bra{0} + \hat{\rho}_1(t)\ket{1}\bra{1}$ at some time $t = t_0$, then it remains to be in this form for all time $t$. If we start from a general density operator $\hat{\rho}_{s}(t)$, then analyzing the time-evolution of off-diagonal terms (i.e., $\hat{\rho}_{0,1}(t)$ and $\hat{\rho}_{1,0}(t)$) would be challenging and it could be our next stage of research. 
For \lb{} equation, the same phenomenon appears.
This approach of only considering diagonal operators with vanishing $\hat{\rho}_{0,1}(t)$ and $\hat{\rho}_{1,0}(t)$, has its restriction but it could lead into simple equations after applying Wigner transformation. The semi-classical limits for \rf{} equation and \lb{} equation have similar form.

Some continuing works could include (1) how to perform further
approximations to simplify the hopping coefficients in LCME to obtain
a simpler equation; (2) a systematic and controllable numerical method
to solve \lb{} equation in infinite dimensional Hilbert space
$L^2(\Real)\otimes \text{span}\bigl\{\ket{0}, \ket{1}\bigr\}$. We
will leave these for future works.

\bibliography{ah}

\begin{thebibliography}{36}%
\makeatletter
\providecommand \@ifxundefined [1]{%
 \@ifx{#1\undefined}
}%
\providecommand \@ifnum [1]{%
 \ifnum #1\expandafter \@firstoftwo
 \else \expandafter \@secondoftwo
 \fi
}%
\providecommand \@ifx [1]{%
 \ifx #1\expandafter \@firstoftwo
 \else \expandafter \@secondoftwo
 \fi
}%
\providecommand \natexlab [1]{#1}%
\providecommand \enquote  [1]{``#1''}%
\providecommand \bibnamefont  [1]{#1}%
\providecommand \bibfnamefont [1]{#1}%
\providecommand \citenamefont [1]{#1}%
\providecommand \href@noop [0]{\@secondoftwo}%
\providecommand \href [0]{\begingroup \@sanitize@url \@href}%
\providecommand \@href[1]{\@@startlink{#1}\@@href}%
\providecommand \@@href[1]{\endgroup#1\@@endlink}%
\providecommand \@sanitize@url [0]{\catcode `\\12\catcode `\$12\catcode
  `\&12\catcode `\#12\catcode `\^12\catcode `\_12\catcode `\%12\relax}%
\providecommand \@@startlink[1]{}%
\providecommand \@@endlink[0]{}%
\providecommand \url  [0]{\begingroup\@sanitize@url \@url }%
\providecommand \@url [1]{\endgroup\@href {#1}{\urlprefix }}%
\providecommand \urlprefix  [0]{URL }%
\providecommand \Eprint [0]{\href }%
\providecommand \doibase [0]{http://dx.doi.org/}%
\providecommand \selectlanguage [0]{\@gobble}%
\providecommand \bibinfo  [0]{\@secondoftwo}%
\providecommand \bibfield  [0]{\@secondoftwo}%
\providecommand \translation [1]{[#1]}%
\providecommand \BibitemOpen [0]{}%
\providecommand \bibitemStop [0]{}%
\providecommand \bibitemNoStop [0]{.\EOS\space}%
\providecommand \EOS [0]{\spacefactor3000\relax}%
\providecommand \BibitemShut  [1]{\csname bibitem#1\endcsname}%
\let\auto@bib@innerbib\@empty
\bibitem [{\citenamefont {Ghosh}\ \emph {et~al.}(2004)\citenamefont {Ghosh},
  \citenamefont {Damle}, \citenamefont {Datta},\ and\ \citenamefont
  {Nitzan}}]{Ghosh04}%
  \BibitemOpen
  \bibfield  {author} {\bibinfo {author} {\bibfnamefont {A.}~\bibnamefont
  {Ghosh}}, \bibinfo {author} {\bibfnamefont {P.}~\bibnamefont {Damle}},
  \bibinfo {author} {\bibfnamefont {S.}~\bibnamefont {Datta}}, \ and\ \bibinfo
  {author} {\bibfnamefont {A.}~\bibnamefont {Nitzan}},\ }\href {\doibase
  10.1557/mrs2004.121} {\bibfield  {journal} {\bibinfo  {journal} {MRS
  Bulletin}\ }\textbf {\bibinfo {volume} {29}},\ \bibinfo {pages} {391}
  (\bibinfo {year} {2004})}\BibitemShut {NoStop}%
\bibitem [{\citenamefont {Galperin}\ \emph
  {et~al.}(2007{\natexlab{a}})\citenamefont {Galperin}, \citenamefont
  {Ratner},\ and\ \citenamefont {Nitzan}}]{Galperin07}%
  \BibitemOpen
  \bibfield  {author} {\bibinfo {author} {\bibfnamefont {M.}~\bibnamefont
  {Galperin}}, \bibinfo {author} {\bibfnamefont {M.~A.}\ \bibnamefont
  {Ratner}}, \ and\ \bibinfo {author} {\bibfnamefont {A.}~\bibnamefont
  {Nitzan}},\ }\href {http://stacks.iop.org/0953-8984/19/i=10/a=103201}
  {\bibfield  {journal} {\bibinfo  {journal} {Journal of Physics: Condensed
  Matter}\ }\textbf {\bibinfo {volume} {19}},\ \bibinfo {pages} {103201}
  (\bibinfo {year} {2007}{\natexlab{a}})}\BibitemShut {NoStop}%
\bibitem [{\citenamefont {Galperin}\ \emph {et~al.}(2008)\citenamefont
  {Galperin}, \citenamefont {Ratner}, \citenamefont {Nitzan},\ and\
  \citenamefont {Troisi}}]{Galperin08}%
  \BibitemOpen
  \bibfield  {author} {\bibinfo {author} {\bibfnamefont {M.}~\bibnamefont
  {Galperin}}, \bibinfo {author} {\bibfnamefont {M.~A.}\ \bibnamefont
  {Ratner}}, \bibinfo {author} {\bibfnamefont {A.}~\bibnamefont {Nitzan}}, \
  and\ \bibinfo {author} {\bibfnamefont {A.}~\bibnamefont {Troisi}},\ }\href
  {\doibase 10.1126/science.1146556} {\bibfield  {journal} {\bibinfo  {journal}
  {Science}\ }\textbf {\bibinfo {volume} {319}},\ \bibinfo {pages} {1056}
  (\bibinfo {year} {2008})},\ \Eprint
  {http://arxiv.org/abs/http://science.sciencemag.org/content/319/5866/1056.full.pdf}
  {http://science.sciencemag.org/content/319/5866/1056.full.pdf} \BibitemShut
  {NoStop}%
\bibitem [{\citenamefont {Holstein}(1959)}]{holstein1959}%
  \BibitemOpen
  \bibfield  {author} {\bibinfo {author} {\bibfnamefont {T.}~\bibnamefont
  {Holstein}},\ }\href {\doibase
  http://dx.doi.org/10.1016/0003-4916(59)90002-8} {\bibfield  {journal}
  {\bibinfo  {journal} {Annals of Physics}\ }\textbf {\bibinfo {volume} {8}},\
  \bibinfo {pages} {325 } (\bibinfo {year} {1959})}\BibitemShut {NoStop}%
\bibitem [{\citenamefont {Dou}\ \emph {et~al.}(2015{\natexlab{a}})\citenamefont
  {Dou}, \citenamefont {Nitzan},\ and\ \citenamefont
  {Subotnik}}]{Wenjie15_friction}%
  \BibitemOpen
  \bibfield  {author} {\bibinfo {author} {\bibfnamefont {W.}~\bibnamefont
  {Dou}}, \bibinfo {author} {\bibfnamefont {A.}~\bibnamefont {Nitzan}}, \ and\
  \bibinfo {author} {\bibfnamefont {J.~E.}\ \bibnamefont {Subotnik}},\ }\href
  {\doibase http://dx.doi.org/10.1063/1.4927237} {\bibfield  {journal}
  {\bibinfo  {journal} {The Journal of Chemical Physics}\ }\textbf {\bibinfo
  {volume} {143}},\ \bibinfo {pages} {054103} (\bibinfo {year}
  {2015}{\natexlab{a}})}\BibitemShut {NoStop}%
\bibitem [{\citenamefont {Ryndyk}(2016)}]{Dmitry}%
  \BibitemOpen
  \bibfield  {author} {\bibinfo {author} {\bibfnamefont {D.~A.}\ \bibnamefont
  {Ryndyk}},\ }\href@noop {} {\emph {\bibinfo {title} {Theory of Quantum
  Transport at Nanoscale electronic resource] : An Introduction}}}\ (\bibinfo
  {publisher} {Springer International Publishing : Imprint: Springer},\
  \bibinfo {address} {Cham},\ \bibinfo {year} {2016})\BibitemShut {NoStop}%
\bibitem [{\citenamefont {Galperin}\ \emph {et~al.}(2004)\citenamefont
  {Galperin}, \citenamefont {Ratner},\ and\ \citenamefont
  {Nitzan}}]{Galperin04_Green}%
  \BibitemOpen
  \bibfield  {author} {\bibinfo {author} {\bibfnamefont {M.}~\bibnamefont
  {Galperin}}, \bibinfo {author} {\bibfnamefont {M.~A.}\ \bibnamefont
  {Ratner}}, \ and\ \bibinfo {author} {\bibfnamefont {A.}~\bibnamefont
  {Nitzan}},\ }\href {\doibase 10.1063/1.1814076} {\bibfield  {journal}
  {\bibinfo  {journal} {The Journal of Chemical Physics}\ }\textbf {\bibinfo
  {volume} {121}},\ \bibinfo {pages} {11965} (\bibinfo {year} {2004})},\
  \Eprint {http://arxiv.org/abs/http://dx.doi.org/10.1063/1.1814076}
  {http://dx.doi.org/10.1063/1.1814076} \BibitemShut {NoStop}%
\bibitem [{\citenamefont {Galperin}\ \emph {et~al.}(2006)\citenamefont
  {Galperin}, \citenamefont {Nitzan},\ and\ \citenamefont
  {Ratner}}]{Galperin06_EOM}%
  \BibitemOpen
  \bibfield  {author} {\bibinfo {author} {\bibfnamefont {M.}~\bibnamefont
  {Galperin}}, \bibinfo {author} {\bibfnamefont {A.}~\bibnamefont {Nitzan}}, \
  and\ \bibinfo {author} {\bibfnamefont {M.~A.}\ \bibnamefont {Ratner}},\
  }\href {\doibase 10.1103/PhysRevB.73.045314} {\bibfield  {journal} {\bibinfo
  {journal} {Phys. Rev. B}\ }\textbf {\bibinfo {volume} {73}},\ \bibinfo
  {pages} {045314} (\bibinfo {year} {2006})}\BibitemShut {NoStop}%
\bibitem [{\citenamefont {Galperin}\ \emph
  {et~al.}(2007{\natexlab{b}})\citenamefont {Galperin}, \citenamefont
  {Nitzan},\ and\ \citenamefont {Ratner}}]{Galperin07_EOM}%
  \BibitemOpen
  \bibfield  {author} {\bibinfo {author} {\bibfnamefont {M.}~\bibnamefont
  {Galperin}}, \bibinfo {author} {\bibfnamefont {A.}~\bibnamefont {Nitzan}}, \
  and\ \bibinfo {author} {\bibfnamefont {M.~A.}\ \bibnamefont {Ratner}},\
  }\href {\doibase 10.1103/PhysRevB.76.035301} {\bibfield  {journal} {\bibinfo
  {journal} {Phys. Rev. B}\ }\textbf {\bibinfo {volume} {76}},\ \bibinfo
  {pages} {035301} (\bibinfo {year} {2007}{\natexlab{b}})}\BibitemShut
  {NoStop}%
\bibitem [{\citenamefont {M\"uhlbacher}\ and\ \citenamefont
  {Rabani}(2008)}]{Rabani}%
  \BibitemOpen
  \bibfield  {author} {\bibinfo {author} {\bibfnamefont {L.}~\bibnamefont
  {M\"uhlbacher}}\ and\ \bibinfo {author} {\bibfnamefont {E.}~\bibnamefont
  {Rabani}},\ }\href {\doibase 10.1103/PhysRevLett.100.176403} {\bibfield
  {journal} {\bibinfo  {journal} {Phys. Rev. Lett.}\ }\textbf {\bibinfo
  {volume} {100}},\ \bibinfo {pages} {176403} (\bibinfo {year}
  {2008})}\BibitemShut {NoStop}%
\bibitem [{\citenamefont {Mitra}\ \emph {et~al.}(2004)\citenamefont {Mitra},
  \citenamefont {Aleiner},\ and\ \citenamefont {Millis}}]{Mitra04}%
  \BibitemOpen
  \bibfield  {author} {\bibinfo {author} {\bibfnamefont {A.}~\bibnamefont
  {Mitra}}, \bibinfo {author} {\bibfnamefont {I.}~\bibnamefont {Aleiner}}, \
  and\ \bibinfo {author} {\bibfnamefont {A.~J.}\ \bibnamefont {Millis}},\
  }\href {\doibase 10.1103/PhysRevB.69.245302} {\bibfield  {journal} {\bibinfo
  {journal} {Phys. Rev. B}\ }\textbf {\bibinfo {volume} {69}},\ \bibinfo
  {pages} {245302} (\bibinfo {year} {2004})}\BibitemShut {NoStop}%
\bibitem [{\citenamefont {Mitra}\ \emph {et~al.}(2005)\citenamefont {Mitra},
  \citenamefont {Aleiner},\ and\ \citenamefont {Millis}}]{Mitra05}%
  \BibitemOpen
  \bibfield  {author} {\bibinfo {author} {\bibfnamefont {A.}~\bibnamefont
  {Mitra}}, \bibinfo {author} {\bibfnamefont {I.}~\bibnamefont {Aleiner}}, \
  and\ \bibinfo {author} {\bibfnamefont {A.~J.}\ \bibnamefont {Millis}},\
  }\href {\doibase 10.1103/PhysRevLett.94.076404} {\bibfield  {journal}
  {\bibinfo  {journal} {Phys. Rev. Lett.}\ }\textbf {\bibinfo {volume} {94}},\
  \bibinfo {pages} {076404} (\bibinfo {year} {2005})}\BibitemShut {NoStop}%
\bibitem [{\citenamefont {Chen}\ \emph {et~al.}(2016)\citenamefont {Chen},
  \citenamefont {Cohen}, \citenamefont {Millis},\ and\ \citenamefont
  {Reichman}}]{ChenHsingTa2016}%
  \BibitemOpen
  \bibfield  {author} {\bibinfo {author} {\bibfnamefont {H.-T.}\ \bibnamefont
  {Chen}}, \bibinfo {author} {\bibfnamefont {G.}~\bibnamefont {Cohen}},
  \bibinfo {author} {\bibfnamefont {A.~J.}\ \bibnamefont {Millis}}, \ and\
  \bibinfo {author} {\bibfnamefont {D.~R.}\ \bibnamefont {Reichman}},\ }\href
  {\doibase 10.1103/PhysRevB.93.174309} {\bibfield  {journal} {\bibinfo
  {journal} {Phys. Rev. B}\ }\textbf {\bibinfo {volume} {93}},\ \bibinfo
  {pages} {174309} (\bibinfo {year} {2016})}\BibitemShut {NoStop}%
\bibitem [{\citenamefont {Elste}\ \emph {et~al.}(2008)\citenamefont {Elste},
  \citenamefont {Weick}, \citenamefont {Timm},\ and\ \citenamefont {von
  Oppen}}]{Elste2008}%
  \BibitemOpen
  \bibfield  {author} {\bibinfo {author} {\bibfnamefont {F.}~\bibnamefont
  {Elste}}, \bibinfo {author} {\bibfnamefont {G.}~\bibnamefont {Weick}},
  \bibinfo {author} {\bibfnamefont {C.}~\bibnamefont {Timm}}, \ and\ \bibinfo
  {author} {\bibfnamefont {F.}~\bibnamefont {von Oppen}},\ }\href {\doibase
  10.1007/s00339-008-4826-2} {\bibfield  {journal} {\bibinfo  {journal}
  {Applied Physics A}\ }\textbf {\bibinfo {volume} {93}},\ \bibinfo {pages}
  {345} (\bibinfo {year} {2008})}\BibitemShut {NoStop}%
\bibitem [{\citenamefont {Esposito}\ and\ \citenamefont
  {Galperin}(2009)}]{Esposito09}%
  \BibitemOpen
  \bibfield  {author} {\bibinfo {author} {\bibfnamefont {M.}~\bibnamefont
  {Esposito}}\ and\ \bibinfo {author} {\bibfnamefont {M.}~\bibnamefont
  {Galperin}},\ }\href {\doibase 10.1103/PhysRevB.79.205303} {\bibfield
  {journal} {\bibinfo  {journal} {Phys. Rev. B}\ }\textbf {\bibinfo {volume}
  {79}},\ \bibinfo {pages} {205303} (\bibinfo {year} {2009})}\BibitemShut
  {NoStop}%
\bibitem [{\citenamefont {Esposito}\ and\ \citenamefont
  {Galperin}(2010)}]{Esposito10}%
  \BibitemOpen
  \bibfield  {author} {\bibinfo {author} {\bibfnamefont {M.}~\bibnamefont
  {Esposito}}\ and\ \bibinfo {author} {\bibfnamefont {M.}~\bibnamefont
  {Galperin}},\ }\href {\doibase 10.1021/jp103369s} {\bibfield  {journal}
  {\bibinfo  {journal} {The Journal of Physical Chemistry C}\ }\textbf
  {\bibinfo {volume} {114}},\ \bibinfo {pages} {20362} (\bibinfo {year}
  {2010})}\BibitemShut {NoStop}%
\bibitem [{\citenamefont {Dou}\ \emph {et~al.}(2015{\natexlab{b}})\citenamefont
  {Dou}, \citenamefont {Nitzan},\ and\ \citenamefont {Subotnik}}]{Wenjie15_ah}%
  \BibitemOpen
  \bibfield  {author} {\bibinfo {author} {\bibfnamefont {W.}~\bibnamefont
  {Dou}}, \bibinfo {author} {\bibfnamefont {A.}~\bibnamefont {Nitzan}}, \ and\
  \bibinfo {author} {\bibfnamefont {J.~E.}\ \bibnamefont {Subotnik}},\ }\href
  {\doibase http://dx.doi.org/10.1063/1.4908034} {\bibfield  {journal}
  {\bibinfo  {journal} {The Journal of Chemical Physics}\ }\textbf {\bibinfo
  {volume} {142}},\ \bibinfo {pages} {084110} (\bibinfo {year}
  {2015}{\natexlab{b}})}\BibitemShut {NoStop}%
\bibitem [{\citenamefont {Dou}\ \emph {et~al.}(2015{\natexlab{c}})\citenamefont
  {Dou}, \citenamefont {Nitzan},\ and\ \citenamefont
  {Subotnik}}]{Wenjie15_broadening}%
  \BibitemOpen
  \bibfield  {author} {\bibinfo {author} {\bibfnamefont {W.}~\bibnamefont
  {Dou}}, \bibinfo {author} {\bibfnamefont {A.}~\bibnamefont {Nitzan}}, \ and\
  \bibinfo {author} {\bibfnamefont {J.~E.}\ \bibnamefont {Subotnik}},\ }\href
  {\doibase http://dx.doi.org/10.1063/1.4922513} {\bibfield  {journal}
  {\bibinfo  {journal} {The Journal of Chemical Physics}\ }\textbf {\bibinfo
  {volume} {142}},\ \bibinfo {pages} {234106} (\bibinfo {year}
  {2015}{\natexlab{c}})}\BibitemShut {NoStop}%
\bibitem [{\citenamefont {Dou}\ and\ \citenamefont
  {Subotnik}(2016)}]{Wenjie16_bcme}%
  \BibitemOpen
  \bibfield  {author} {\bibinfo {author} {\bibfnamefont {W.}~\bibnamefont
  {Dou}}\ and\ \bibinfo {author} {\bibfnamefont {J.~E.}\ \bibnamefont
  {Subotnik}},\ }\href {\doibase http://dx.doi.org/10.1063/1.4939734}
  {\bibfield  {journal} {\bibinfo  {journal} {The Journal of Chemical Physics}\
  }\textbf {\bibinfo {volume} {144}},\ \bibinfo {pages} {024116} (\bibinfo
  {year} {2016})}\BibitemShut {NoStop}%
\bibitem [{\citenamefont {Breuer}\ and\ \citenamefont
  {Petruccione}(2002)}]{breuer}%
  \BibitemOpen
  \bibfield  {author} {\bibinfo {author} {\bibfnamefont {H.-P.}\ \bibnamefont
  {Breuer}}\ and\ \bibinfo {author} {\bibfnamefont {F.}~\bibnamefont
  {Petruccione}},\ }\href@noop {} {\emph {\bibinfo {title} {The Theory of Open
  Quantum Systems}}}\ (\bibinfo  {publisher} {Oxford University Press},\
  \bibinfo {year} {2002})\BibitemShut {NoStop}%
\bibitem [{\citenamefont {Breuer}\ \emph {et~al.}(2016)\citenamefont {Breuer},
  \citenamefont {Laine}, \citenamefont {Piilo},\ and\ \citenamefont
  {Vacchini}}]{breuer2016}%
  \BibitemOpen
  \bibfield  {author} {\bibinfo {author} {\bibfnamefont {H.-P.}\ \bibnamefont
  {Breuer}}, \bibinfo {author} {\bibfnamefont {E.-M.}\ \bibnamefont {Laine}},
  \bibinfo {author} {\bibfnamefont {J.}~\bibnamefont {Piilo}}, \ and\ \bibinfo
  {author} {\bibfnamefont {B.}~\bibnamefont {Vacchini}},\ }\href {\doibase
  10.1103/RevModPhys.88.021002} {\bibfield  {journal} {\bibinfo  {journal}
  {Rev. Mod. Phys.}\ }\textbf {\bibinfo {volume} {88}},\ \bibinfo {pages}
  {021002} (\bibinfo {year} {2016})}\BibitemShut {NoStop}%
\bibitem [{\citenamefont {Davies}(1974)}]{Davies}%
  \BibitemOpen
  \bibfield  {author} {\bibinfo {author} {\bibfnamefont {E.~B.}\ \bibnamefont
  {Davies}},\ }\href {http://projecteuclid.org/euclid.cmp/1103860160}
  {\bibfield  {journal} {\bibinfo  {journal} {Comm. Math. Phys.}\ }\textbf
  {\bibinfo {volume} {39}},\ \bibinfo {pages} {91} (\bibinfo {year}
  {1974})}\BibitemShut {NoStop}%
\bibitem [{\citenamefont {Lindblad}(1976)}]{lindblad1976}%
  \BibitemOpen
  \bibfield  {author} {\bibinfo {author} {\bibfnamefont {G.}~\bibnamefont
  {Lindblad}},\ }\href@noop {} {\bibfield  {journal} {\bibinfo  {journal}
  {Comm. Math. Phys.}\ }\textbf {\bibinfo {volume} {48}},\ \bibinfo {pages}
  {119} (\bibinfo {year} {1976})}\BibitemShut {NoStop}%
\bibitem [{\citenamefont {McCracken}(2013)}]{McCracken2013}%
  \BibitemOpen
  \bibfield  {author} {\bibinfo {author} {\bibfnamefont {J.~M.}\ \bibnamefont
  {McCracken}},\ }\href {\doibase 10.1103/PhysRevA.88.032103} {\bibfield
  {journal} {\bibinfo  {journal} {Phys. Rev. A}\ }\textbf {\bibinfo {volume}
  {88}},\ \bibinfo {pages} {032103} (\bibinfo {year} {2013})}\BibitemShut
  {NoStop}%
\bibitem [{\citenamefont {Argentieri}\ \emph {et~al.}(2014)\citenamefont
  {Argentieri}, \citenamefont {Benatti}, \citenamefont {Floreanini},\ and\
  \citenamefont {Pezzutto}}]{argentieri}%
  \BibitemOpen
  \bibfield  {author} {\bibinfo {author} {\bibfnamefont {G.}~\bibnamefont
  {Argentieri}}, \bibinfo {author} {\bibfnamefont {F.}~\bibnamefont {Benatti}},
  \bibinfo {author} {\bibfnamefont {R.}~\bibnamefont {Floreanini}}, \ and\
  \bibinfo {author} {\bibfnamefont {M.}~\bibnamefont {Pezzutto}},\ }\href
  {http://stacks.iop.org/0295-5075/107/i=5/a=50007} {\bibfield  {journal}
  {\bibinfo  {journal} {EPL (Europhysics Letters)}\ }\textbf {\bibinfo {volume}
  {107}},\ \bibinfo {pages} {50007} (\bibinfo {year} {2014})}\BibitemShut
  {NoStop}%
\bibitem [{\citenamefont {Pechukas}(1994)}]{Pechukas1994}%
  \BibitemOpen
  \bibfield  {author} {\bibinfo {author} {\bibfnamefont {P.}~\bibnamefont
  {Pechukas}},\ }\href {\doibase 10.1103/PhysRevLett.73.1060} {\bibfield
  {journal} {\bibinfo  {journal} {Phys. Rev. Lett.}\ }\textbf {\bibinfo
  {volume} {73}},\ \bibinfo {pages} {1060} (\bibinfo {year}
  {1994})}\BibitemShut {NoStop}%
\bibitem [{\citenamefont {Shaji}\ and\ \citenamefont
  {Sudarshan}(2005)}]{Shaji200548}%
  \BibitemOpen
  \bibfield  {author} {\bibinfo {author} {\bibfnamefont {A.}~\bibnamefont
  {Shaji}}\ and\ \bibinfo {author} {\bibfnamefont {E.}~\bibnamefont
  {Sudarshan}},\ }\href {\doibase
  http://dx.doi.org/10.1016/j.physleta.2005.04.029} {\bibfield  {journal}
  {\bibinfo  {journal} {Physics Letters A}\ }\textbf {\bibinfo {volume}
  {341}},\ \bibinfo {pages} {48 } (\bibinfo {year} {2005})}\BibitemShut
  {NoStop}%
\bibitem [{\citenamefont {Shankar}(1994)}]{shankar}%
  \BibitemOpen
  \bibfield  {author} {\bibinfo {author} {\bibfnamefont {R.}~\bibnamefont
  {Shankar}},\ }\href@noop {} {\emph {\bibinfo {title} {Principles of Quantum
  Mechanics}}}\ (\bibinfo  {publisher} {Springer},\ \bibinfo {year}
  {1994})\BibitemShut {NoStop}%
\bibitem [{\citenamefont {Blum}(1996)}]{densitymat}%
  \BibitemOpen
  \bibfield  {author} {\bibinfo {author} {\bibfnamefont {K.}~\bibnamefont
  {Blum}},\ }\href@noop {} {\emph {\bibinfo {title} {Density matrix theory and
  applications}}}\ (\bibinfo  {publisher} {New York : Plenum Press},\ \bibinfo
  {year} {1996})\BibitemShut {NoStop}%
\bibitem [{\citenamefont {Bruch}\ \emph {et~al.}(2016)\citenamefont {Bruch},
  \citenamefont {Thomas}, \citenamefont {Viola~Kusminskiy}, \citenamefont {von
  Oppen},\ and\ \citenamefont {Nitzan}}]{Bruch2016}%
  \BibitemOpen
  \bibfield  {author} {\bibinfo {author} {\bibfnamefont {A.}~\bibnamefont
  {Bruch}}, \bibinfo {author} {\bibfnamefont {M.}~\bibnamefont {Thomas}},
  \bibinfo {author} {\bibfnamefont {S.}~\bibnamefont {Viola~Kusminskiy}},
  \bibinfo {author} {\bibfnamefont {F.}~\bibnamefont {von Oppen}}, \ and\
  \bibinfo {author} {\bibfnamefont {A.}~\bibnamefont {Nitzan}},\ }\href
  {\doibase 10.1103/PhysRevB.93.115318} {\bibfield  {journal} {\bibinfo
  {journal} {Phys. Rev. B}\ }\textbf {\bibinfo {volume} {93}},\ \bibinfo
  {pages} {115318} (\bibinfo {year} {2016})}\BibitemShut {NoStop}%
\bibitem [{\citenamefont {Ouyang}\ \emph {et~al.}(2015)\citenamefont {Ouyang},
  \citenamefont {Dou},\ and\ \citenamefont {Subotnik}}]{Wenjie15_leaking}%
  \BibitemOpen
  \bibfield  {author} {\bibinfo {author} {\bibfnamefont {W.}~\bibnamefont
  {Ouyang}}, \bibinfo {author} {\bibfnamefont {W.}~\bibnamefont {Dou}}, \ and\
  \bibinfo {author} {\bibfnamefont {J.~E.}\ \bibnamefont {Subotnik}},\ }\href
  {\doibase http://dx.doi.org/10.1063/1.4908032} {\bibfield  {journal}
  {\bibinfo  {journal} {The Journal of Chemical Physics}\ }\textbf {\bibinfo
  {volume} {142}},\ \bibinfo {pages} {084109} (\bibinfo {year}
  {2015})}\BibitemShut {NoStop}%
\bibitem [{\citenamefont {Wigner}(1932)}]{wigner1932}%
  \BibitemOpen
  \bibfield  {author} {\bibinfo {author} {\bibfnamefont {E.}~\bibnamefont
  {Wigner}},\ }\href {\doibase 10.1103/PhysRev.40.749} {\bibfield  {journal}
  {\bibinfo  {journal} {Phys. Rev.}\ }\textbf {\bibinfo {volume} {40}},\
  \bibinfo {pages} {749} (\bibinfo {year} {1932})}\BibitemShut {NoStop}%
\bibitem [{\citenamefont {Zworski}(2012)}]{Zworski}%
  \BibitemOpen
  \bibfield  {author} {\bibinfo {author} {\bibfnamefont {M.}~\bibnamefont
  {Zworski}},\ }\href@noop {} {\emph {\bibinfo {title} {Semiclassical
  Analysis}}}\ (\bibinfo  {publisher} {American Mathematical Society},\
  \bibinfo {year} {2012})\ \bibinfo {note} {graduate Studies in Mathematics,
  Volume 138}\BibitemShut {NoStop}%
\bibitem [{Note1()}]{Note1}%
  \BibitemOpen
  \bibinfo {note} {We choose to follow the usual convention here, which differs
  by a negative sign compared to \cite {Zworski}}\BibitemShut {NoStop}%
\bibitem [{\citenamefont {Curtright}\ \emph {et~al.}(2014)\citenamefont
  {Curtright}, \citenamefont {Fairlie},\ and\ \citenamefont
  {Zachos}}]{curtright}%
  \BibitemOpen
  \bibfield  {author} {\bibinfo {author} {\bibfnamefont {T.~L.}\ \bibnamefont
  {Curtright}}, \bibinfo {author} {\bibfnamefont {D.~B.}\ \bibnamefont
  {Fairlie}}, \ and\ \bibinfo {author} {\bibfnamefont {C.~K.}\ \bibnamefont
  {Zachos}},\ }\href@noop {} {\emph {\bibinfo {title} {A Concise Treatise on
  Quantum Mechanics in Phase Space}}}\ (\bibinfo  {publisher} {World Scientific
  Publishing Co. Pte. Ltd},\ \bibinfo {year} {2014})\BibitemShut {NoStop}%
\bibitem [{\citenamefont {Koch}\ and\ \citenamefont {von
  Oppen}(2005)}]{Jens2005}%
  \BibitemOpen
  \bibfield  {author} {\bibinfo {author} {\bibfnamefont {J.}~\bibnamefont
  {Koch}}\ and\ \bibinfo {author} {\bibfnamefont {F.}~\bibnamefont {von
  Oppen}},\ }\href {\doibase 10.1103/PhysRevLett.94.206804} {\bibfield
  {journal} {\bibinfo  {journal} {Phys. Rev. Lett.}\ }\textbf {\bibinfo
  {volume} {94}},\ \bibinfo {pages} {206804} (\bibinfo {year}
  {2005})}\BibitemShut {NoStop}%
\end{thebibliography}%

\end{document}